\documentclass[12pt,letterpaper]{article}

\usepackage[includeheadfoot,
            marginratio={1:1,2:3}, 
            width=412pt, 
            height=688pt,]{geometry}

\usepackage{amsmath}
\usepackage{amsfonts}
\usepackage{amssymb}
\usepackage{graphicx}
\usepackage{cite}
\usepackage{ulem}
\usepackage{dsfont}
\usepackage{longtable}
\usepackage{afterpage}
\usepackage{booktabs}
\usepackage{hyperref}
\usepackage{tikz} 
\usepackage{subfigure}

\newcommand{\eq}[1]{\begin{equation}
                     \begin{split} #1 \end{split}
                     \end{equation}}

\newcommand{\ov}{\overline}
\newcommand{\op}{\hspace{1pt}}
\newcommand{\parag}{\,\begin{matrix} \gtrsim \\[-0.3cm]{}_p \end{matrix}\,}
\newcommand{\paraeq}{\,\begin{matrix} \simeq \\[-0.3cm]{}_p \end{matrix}\,}
\newcommand{\paral}{\,\begin{matrix} \lesssim \\[-0.3cm]{}_p \end{matrix}\,}

\allowdisplaybreaks[2]
\numberwithin{equation}{section}

\begin{document}

\normalem
\vspace*{-1.5cm}
\begin{flushright}
  {\small
  MPP-2017-34 \\
  }
\end{flushright}

\vspace{1.5cm}

\begin{center}
  {\LARGE
    The Swampland Conjecture and   F-term \\[0.3cm]
    Axion Monodromy Inflation 
}
\vspace{0.4cm}

\end{center}

\vspace{0.35cm}
\begin{center}
  Ralph Blumenhagen$^1$, Irene Valenzuela$^{1,2}$, Florian Wolf$\,^1$
\end{center}

\vspace{0.1cm}
\begin{center} 
\emph{$^1$
Max-Planck-Institut f\"ur Physik (Werner-Heisenberg-Institut), \\ 
F\"ohringer Ring 6,  80805 M\"unchen, Germany 
} 

\vspace{0.2cm}
\emph{$^2$
Institute for Theoretical Physics and Center for Extreme Matter and \\Emergent Phenomena,
Utrecht University, \\ Princetonplein 5, 3584 CC Utrecht, The Netherlands
}

\end{center} 

\vspace{1cm}


\begin{abstract}
\noindent
We continue the investigation of F-term axion monodromy
inflation in string theory, while seriously taking the 
issue of moduli stabilization into account. 
For a number of closed and open string models,
we show that they suffer from  serious control issues once one is trying to realize
trans-Planckian field excursions. More precisely, the flux tuning required to delay the logarithmic scaling of the field distance to a trans-Planckian value cannot be done without leaving the regime where the employed  effective
 supergravity theory is under control. 
Our findings are consistent with the axionic extension of the Refined
Swampland Conjecture, 
stating that in quantum gravity
the effective theory breaks down for a field excursion beyond  the Planck scale.
Our analysis suggests that models of F-term axion monodromy inflation
 with a tensor-to-scalar ratio $r\ge O(10^{-3})$ cannot
be parametrically  controlled.
\end{abstract}


\clearpage
\tableofcontents


\section{Introduction}

Even though meanwhile dismissed, the 2014 BICEP2 announcement of a 
detection of primordial B-modes with a large tensor-to-scalar ratio of
$r \sim 0.2$, triggered much research in string cosmology. 
Indeed, the main model building challenge is that for a ratio of $r>
0.01$   the Lyth bound \cite{Lyth:1996im} implies 
that the inflaton has to roll   over trans-Planckian field distances, 
hence  making the process highly UV sensitive. 
Therefore, string theory as a UV complete quantum theory of gravity
provides a well defined framework to discuss high scale inflation. Interestingly, there are some hints supporting the existence of an underlying quantum gravity constraint that forbids trans-Planckian excursions. Further investigation in this direction is therefore, not only phenomenological, but also conceptually interesting.

To forbid higher order Planck suppressed operators in the inflaton action,
one can employ  a pseudo-scalar field with a continuous shift symmetry, called an axion. 
There are essentially two mostly followed approaches  towards
realizing  axionic inflation in string theory. 
The first employs  the periodic cosine potential \cite{Freese:1990rb}
generically generated by instantons, possibly with more than one axion
to enlarge the field range \cite{Kim:2004rp,Dimopoulos:2005ac}. 
For the simplest model of natural inflation, string theory requires
to work outside the regime of a controlled low-energy effective action
\cite{Svrcek:2006yi}. It was realized
\cite{Rudelius:2014wla, Rudelius:2015xta, Montero:2015ofa,Brown:2015iha} 
that this behavior is precisely reflected in
the Weak Gravity Conjecture (WGC) \cite{ArkaniHamed:2006dz} extended from
point particles to instantons.

The second approach is to impose  a controlled spontaneous breaking of the
axionic shift symmetry \cite{Kaloper:2008fb} by adding branes or
fluxes, inducing  a potential energy  that increases by 
a certain amount over every period the inflaton transverses.
This ansatz is called axion monodromy inflation and was introduced in
the stringy context in \cite{Silverstein:2008sg}. One mechanism  to
generate  a polynomial potential for  axion monodromy inflation 
is to turn on background fluxes generating a tree-level F-term scalar potential \cite{Marchesano:2014mla,Blumenhagen:2014gta,Hebecker:2014eua},
 see also \cite{Palti:2014kza,Grimm:2014vva,Ibanez:2014kia,Arends:2014qca,Hassler:2014mla,McAllister:2014mpa,Ibanez:2014swa,Buchmuller:2015oma,Retolaza:2015sta,Bielleman:2015ina} and for reviews \cite{Baumann:2014nda,Westphal:2014ana}. 
For other attempts to realize axion monodromy inflation in string
theory see e.g. \cite{Hebecker:2014kva,Escobar:2015fda,Escobar:2015ckf}.

Turning on fluxes has the advantage that the same mechanism
generating the axion potential also stabilizes the other moduli and
breaks supersymmetry. Therefore, the question arose whether 
one can control the trans-Planckian regime for the axion 
in a consistent scheme of moduli stabilization.
This was analyzed in a series of papers \cite{Blumenhagen:2014nba, Blumenhagen:2015kja,Blumenhagen:2015qda,Blumenhagen:2015xpa}
in the framework of orientifolded Calabi-Yau compactification of
the 10-dimensional type IIA or type IIB theory giving rise to a four
dimensional $N = 1$ supergravity theory with usually plenty of
massless scalar fields and axions. This
geometry is then perturbed by turning on geometric and non-geometric
background fluxes leading to a gauged supergravity theory, that can be deduced via dimensional
reduction of double field theory \cite{Blumenhagen:2015lta}. 

A detectable tensor-to-scalar ratio of $r>0.01$ and the so far not
detected non-Gaussianities  favor  single large-field inflation. 
In this case, the potential energy  during inflation is $M_{\rm inf}
\sim 10^{16}$ GeV, the Hubble-scale 
during inflation is $H_{\rm inf} \sim 10^{14}$ GeV and the inflaton
mass is $M_{\theta} \sim 10^{13}$ GeV. 
In order to use an effective supergravity approach, 
the string scale $M_{\rm s}$ and the Kaluza-Klein scale $M_{\rm KK}$ must lie
above all these scales.
Moreover,  the other moduli masses should lie above the Hubble scale to guarantee a model of single field inflation. Therefore, altogether we have the ordered hierarchy of mass scales
\eq{
\label{introhierarchy}
M_{\rm Pl} \; > \; M_{\rm s} \; > \; M_{\rm KK} \; > \;M_{\rm mod} \; > \; H_{\rm inf} \; > \; M_{\rm \theta } \,,
}
where neighboring scales can differ only by a factor of ${\mathcal
  O}(10)$. This is obviously a major challenge for concrete string
model building.

Since for single field
inflation,  the inflaton should be the lightest scalar field, all other
moduli should better acquire their masses already at tree-level.
In the type IIB setting this implies that the universal axio-dilaton
requires an NS-NS three-form flux and the overall volume a
non-geometric $Q$-flux to be turned on.
Closed string moduli stabilization with solely
fluxes was discussed in
\cite{Blumenhagen:2015kja,Blumenhagen:2015qda} (see also \cite{Hebecker:2014kva}). 
There it was found that control over the trans-Planckian regime
in all examples required to violate at least one of the required
hierarchies in \eqref{introhierarchy}. Moreover, the backreaction of the rolling
axion onto the other moduli was substantial and led to a flattening
of the potential \cite{Dong:2010in} and in the extreme case to a potential of plateau(Starobinsky)-type.
The reason behind this is that for large field excursions of an axion
$\theta$, the backreacted proper field distance showed a logarithmic behavior 
$\Theta\sim \lambda^{-1} \log\theta$.
Here $\lambda^{-1}$ can be considered as the scale in field distance where the backreaction
becomes substantial.

It was realized in \cite{Baume:2016psm,Klaewer:2016kiy} that this logarithmic scaling of the proper
field distance is very generic and that it  precisely reflects  the
conjectured behavior  by  Ooguri/Vafa \cite{Ooguri:2006in} to distinguish 
effective field theory models that can be realized in string theory
(the landscape) from those that cannot be coupled in a UV complete way
to gravity (the swampland) \cite{Vafa:2005ui}.
This, later called, swampland conjecture \cite{Klaewer:2016kiy} 
says that if one moves over very large distances in the moduli
space  of  an effective  quantum gravity theory, there appears an 
infinite tower of states whose mass scales as $m\sim m_0 \exp(-\lambda
\Delta\Theta)$. This means that for $\Delta\Theta> \lambda^{-1}$ 
the effective theory breaks down. 
The prototype example of this appears for string theory compactified
on a circle, where it is the Kaluza-Klein tower that shows this
behavior in terms of the proper field distance.

The string theory models discussed in \cite{Baume:2016psm} always had
$\lambda=O(1)$, i.e. the cut-off in the field distance where one
could trust the effective description was close to the Planck-scale.
This led Kl\"awer and Palti in \cite{Klaewer:2016kiy}, to formulate the {\it Refined Swampland
  Conjecture} (RSC), extending the former one by the statement that
$\lambda=O(1)$, i.e. one cannot push $\lambda^{-1}$ 
to values parametrically larger than one. Furthermore, the RSC applies to any scalar field, including axions, unlike the original conjecture from  \cite{Ooguri:2006in} which only applies to the geometric moduli space.

It was motivated in \cite{Valenzuela:2016yny}, though, that one should aim for engineering models with a flux dependent $\lambda$ in such a way that the backreaction can in principle be delayed in field distance. The authors of \cite{Bielleman:2016olv}
analyzed inflationary models with an open string modulus, namely
the deformation modulus of a $D7$-brane \cite{Hebecker:2014eua,Arends:2014qca,Ibanez:2014kia,Ibanez:2014swa,Bielleman:2015lka,Bielleman:2016grv}, playing the role
of the inflaton. These models looked a priori promising to admit
a parametrically large value of $\lambda^{-1}$.
However, the {\it Refined Swampland
  Conjecture} implies that also F-term axion monodromy inflation
cannot be realized in a parametrically controlled way in string
theory. Let us mention  that an argument based on entropy of de-Sitter space
has led  J.~Conlon to the same general conclusion \cite{Conlon:2012tz}
(see also \cite{Kaloper:2015jcz}).

It is the purpose of this paper to challenge or find further evidence
for this intricate  relation between F-term axion monodromy inflation
and the {\it Refined Swampland Conjecture}.
Despite the danger of repeating parts of this introduction,
in section 2 we review former attempts to build string models
of large field inflation, discuss the challenges one faces when
combining this with full moduli stabilization and 
 also present  the swampland conjecture and its refinement.
In section 3, we will revisite a simple purely closed string model from \cite{Blumenhagen:2015kja,Blumenhagen:2015qda}
and demonstrate how it fits nicely into this picture.
Moreover, we will show that also the proposed backreacted plateau-like
model \cite{Blumenhagen:2015kja} is not parametrically under control.

In section 4, we extend  and further examine the open string models discussed in \cite{Bielleman:2016olv}. We indeed find that the backreacted proper field distance always exhibit the predicted logarithmic scaling at large field. Our aim is, though, to identify and analyze in detail models where $\lambda^{-1}$  is flux-dependent and can in principle be tuned parametrically large to delay the backreaction. We find that also these models require $\lambda\approx 1$ in order
to have  parametric control over the effective field theories. For concreteness, we consider models in which all scalars are fixed at tree level by fluxes. This requires the addition of geometric fluxes in IIA, which become non-geometric in IIB. We identify two
simple representative models of having a tunable $\lambda$, and show that the necessary flux tuning would imply that the scale of moduli masses becomes larger than the Kaluza-Klein scale.
The (quantum gravity) ingredients in  the string effective action that are responsible
for this behavior can be identified as:
\begin{itemize}
\item{The leading order K\"ahler potential always shows a 
     logarithmic dependence on the saxions.}
\item{The specific form of the superpotential appearing in string theory.}
\item{The moduli dependence of the various mass scales, like string,
    Kaluza-Klein and moduli mass, resulting
    from dimensional reduction and moduli stabilization.}
\item{The fact that fluxes are quantized.}
\end{itemize}

These observations lead us to a change of perspective. Instead
of trying to make the models more baroque and
to find loop-holes, maybe one should better believe in the
{\it Refined Swampland Conjecture} and figure out 
where these control issues were hidden or ignored
in the previous attempts that (naively) looked successful 
to realize large field inflation. 
 We also
critically revisite attempts to build axion monodromy models
where the K\"ahler moduli were stabilized via non-perturbative
effects, like in KKLT and the Large Volume Scenario. We notice that the required flux tuning gets into conflict with the original assumptions of small $W_0$ and large volume, respectively.
Our conclusions in section 5 will also discuss possible loopholes and future directions to continue investigating the realization of axion monodromy inflation and its relation with the Swampland Conjecture.

\section{F-term axion monodromy inflation}
\label{sec_two}

In this section we review former attempts to realize large field
inflation in string theory and challenges one faces,
when combining this with the issue of moduli stabilization.
We also review the Swampland Conjecture \cite{Ooguri:2006in} as
formulated by Ooguri/Vafa
and following \cite{Klaewer:2016kiy} how it is related to large field inflation.

\subsection{Large field inflation}

The large number of difficulties encountered when embedding large field
inflation in a controlled string theory framework gave rise to the
suspicion  that a fundamental reason might underly the obstruction of
getting trans-Planckian field ranges in a consistent theory of quantum
gravity.  The search of this fundamental reason has triggered
plenty of recent work aiming to identify the constraints that quantum
gravity imposes over an, a priori, consistent quantum field 
theory.

The obstruction of getting a trans-Planckian decay constant to realize
natural inflation can be related, for instance, to the Weak Gravity
Conjecture (WGC) \cite{ArkaniHamed:2006dz}. This conjecture
generalized  to axions reads 
\eq{
  f\,S_{\rm inst}\leq 1 \, ,
} 
where $f$ is the  axion decay constant and $S_{\rm inst}$ the instanton action.
Thus, it states that for any axion with a trans-Planckian
decay constant there must exist an instanton, electrically coupled to
the axion, with an action at most of order one. Therefore, the potential for
the axion will generically receive non-suppressed instantonic
corrections which signal the breakdown of the effective theory and
will reduce the effective field range to a sub-Planckian value
\cite{Rudelius:2014wla, Rudelius:2015xta,
  Montero:2015ofa,Brown:2015iha}. Attempts to engineer
trans-Planckian flat directions by using multiple fields are also
highly constrained by strong versions of the Weak Gravity 
Conjecture \cite{Heidenreich:2015wga}.

As outlined in the introduction, a promising alternative is F-term axion monodromy inflation
\cite{Marchesano:2014mla}. The basic idea is to induce a non-periodic
potential for the axion while leaving the discrete shift symmetry
unbroken. This leads to the familiar multi-branched structure which
allows for a non-compact field range for the axion. By rolling down
one of the branches a trans-Planckian excursion can be achieved even if
the axionic decay constant $f$ (and therefore the underlying
periodicity of the system) is sub-Planckian. This implies that the
above constraints coming from the WGC do not apply in this
case. Furthermore, the discrete shift of the axion if combined with a
shift of the integer labeling the different branches is still a
symmetry of the theory. This  protects the effective theory from
dangerous UV corrections coming from states above the cut-off 
scale. 

The realization in four dimensions is given by coupling the axion $\phi$ to a 3-form gauge field $F_4=dC_3$ as follows, 
\eq{\mathcal{L}=-f^2(d\phi)^2- F_4\wedge *F_4+2F_4\,\phi \,.
}
This description was first analyzed in detail by Dvali \cite{Dvali:2005an,Dvali:2005zk} and applied to inflation by Kaloper and Sorbo \cite{Kaloper:2008fb,Kaloper:2011jz,Kaloper:2014zba,Kaloper:2016fbr}.
The gauge field has no
dynamics in four dimensions but its field strength can have a
non-vanishing (quantized) value $f_0$ in the vacuum. Upon integrating out the 3-form field,
\eq{
*F_4=f_0+m\phi\rightarrow V=(f_0+m\phi)^2
}
one recovers the scalar potential for the axion with multiple branches
labeled by $f_0$. Notice that this is not a particular model of
F-term axion monodromy, but a dual formulation in four dimensions,
since for any massive axion one can always define an effective 3-form
field generating the corresponding scalar potential. This formulation
makes the underlying symmetries of the system manifest. In particular,
the combined discrete shift
\eq{
f_0\rightarrow f_0 + c\ ,\qquad \phi\rightarrow \phi -c/m
}
is still a symmetry of the system, and for $c/m=2\pi f$ this
transformation identifies gauge equivalent branches.

Furthermore, transitions between different branches are mediated by
nucleation of membranes electrically charged under the 3-form gauge
field. By crossing a membrane, $f_0$ shifts by an integer times the
charge of the membrane. The tunneling rate is exponentially
suppressed, and can  indeed be estimated by applying the WGC to the
3-form gauge field. However, recent results show that the tunneling
rate is not fast enough to constrain large field 
inflation \cite{Ibanez:2015fcv,Hebecker:2015zss,Brown:2016nqt}. 

Remarkably, this is also the mechanism underlying flux stabilization of
axions in string theory, since the discrete axionic shift symmetry is
indeed a gauge identification and cannot be explicitly broken. As explained, this
does not prevent the axions to become massive in a consistent way with
the discrete shift symmetry. Thus, all axions arising in string
compactifications which are stabilized by internal fluxes are examples
of the aforementioned multi-branched structure and candidates for
F-term axion monodromy. In those cases, the 3-form fields come from
dimensionally reducing higher NS-NS and R-R $p$-form fields and are dual to
the internal  fluxes \cite{Bielleman:2015ina,Carta:2016ynn}.

Despite all these appealing features, including the apparent
robustness against the WGC, we think that there does not exist  any completely successful and
convincing string realization of F-term axion monodromy inflation,
yet. The difficulties are related to moduli stabilization and
backreaction effects from the other scalars of the
compactification\footnote{From this perspective, inflationary string model building
  attempts
  that did not consider  these issues are not yet complete and need to
  be reevaluated.}. When taking the backreaction into account, the
physical field range of the inflaton might be drastically reduced, as
we proceed to explain in section \ref{sec:moduli}. 
 More than a technical issue, these difficulties might again point towards a
fundamental obstruction of any consistent theory of quantum gravity.
As noticed in \cite{Baume:2016psm,Klaewer:2016kiy},
in this case these control issues can be related to the Swampland Conjecture. 
We will review and extend this relation in section \ref{sec_swamp}.

\subsection{Challenge with moduli stabilization}
\label{sec:moduli}

Any attempt to construct a realistic inflationary model in string
theory has to deal with the issue of moduli stabilization. The strong
experimental bounds on non-Gaussianities and isocurvature
perturbations favor a scheme of single field inflation or, at most,
moderate multi-field inflation involving a few weakly-coupled  scalars. 
To guarantee the consistency of the effective field theory
approach as well as to realize a model of single field inflation,
one has to stabilize the moduli such that the following hierarchy of mass
scales is realized 
\eq{
\label{hierarchy1}
M_{\rm Pl} \; > \; M_{\rm s} > M_{\rm KK} > M_{\rm mod} > H_{\rm inf} > M_{\rm \theta} \, ,
}
where $H_{\rm inf}$ is the Hubble scale during inflation and  $ M_{\rm
  \theta}$ the inflaton mass.
These scales are constrained by the
amplitude of scalar density perturbations and the value of the
tensor-to-scalar ratio. For chaotic inflation, $M_\theta\sim 10^{13}$
GeV and $H_{\rm inf}\sim 10^{14}$ GeV. Therefore there is not much
room left to stabilize the rest of the moduli ($M_{\rm mod}$) above
the inflaton mass and below the Kaluza-Klein scale (which is also
usually of order $M_{\rm KK}\sim 10^{16}-10^{17}$ GeV in perturbative
string theory).
To achieve this hierarchy of scales at the minimum of the potential is
already a challenge for many flux compactifications (see
\cite{Blumenhagen:2014nba,Hebecker:2014kva} for some no-go theorems
for the complex structure moduli space of a Calabi-Yau
three-fold). But to guarantee the stabilization of these scales during
the whole inflationary trajectory is an even bigger challenge (see also \cite{Hebecker:2014kva,Buchmuller:2014vda,Dudas:2015lga,Buchmuller:2015oma}).

Let us assume a pseudo-scalar $\theta$ parametrizing the inflationary
trajectory.  When $\theta$ is displaced from its minimum, generically the minima
of the other scalars will  also change,
\eq{
\label{saxion}
s(\theta)=s_0+\delta s(\theta)
}
where $s_0$ denotes the vacuum expectation value of the scalar $s$
at the minimum of the potential, i.e. when $\theta$ is also at its minimum. We will use the word \emph{saxions}
to refer to all non-periodic (non-axionic) scalars. By plugging this
back into the effective theory, the scalar potential and the kinetic
term for the inflaton can be substantially modified. In other words,
the inflationary trajectory is no longer only along $\theta$ but
corresponds to a combination of $\theta$ and $s$. This backreaction
leads to a flattening of the inflaton potential \cite{Dong:2010in}.

Note that the above simple procedure
of freezing $s$ and plugging \eqref{saxion} back into the effective
theory is an approximation that relies on neglecting the variation of
the kinetic energy of the saxion with respect to the potential energy,
so it is valid only as long as there is a mass hierarchy between
$\theta$ and $s$. Otherwise, a multifield analysis is required to
consider simultaneously the dynamics of both  fields.

In the Kaloper-Sorbo formulation of the axion coupled to the 3-form
gauge field, these corrections do not appear from higher dimensional
operators breaking the shift symmetry. They arise from the fact that
the kinetic metric of the 3-form gauge fields is also field dependent
(in particular, it depends on the saxions) \cite{Valenzuela:2016yny}. When integrating out the
3-form gauge field, the shape of the branches becomes field dependent
and can be substantially modified when displacing the inflaton away
from the minimum (in a shift invariant way, but potentially dangerous
for inflation anyway).

In \cite{Blumenhagen:2015qda,Baume:2016psm} it was pointed out that
the displacement of the saxions will generically backreact on the
kinetic metric of the inflaton leading at best to a logarithmic
behavior of the proper field distance at large field. More concretely,
\eq{
\label{hannover96}
\Theta=\int \sqrt{K_{\theta\theta}(s)}\; d\theta\sim \int \frac{1}{s(\theta)}\sim \frac{1}{\lambda} \log(\theta)
}
where we have assumed that $K=-\log(s)$ with $s$ being the saxionic
partner of the inflaton, and that for large field excursions $\delta
s(\theta)\simeq \lambda \theta$. In \eqref{hannover96}, $\Theta$ is
the canonically normalized inflaton field. This implies that
parametrically large displacements are strongly disfavored in string
theory, but in principle trans-Planckian field ranges are still
possible if $\lambda \ll 1$, so that backreaction effects can be
delayed far out in field space. In
other words, the field range available before backreaction effects
become important and the logarithmic scaling takes place, is given by
\eq{\label{thetac}
\Theta_c=\int^{\theta_c} \sqrt{K_{\theta\theta}(s)}\; d\theta\sim \frac{\theta_c}{s_0}\sim \frac{1}{\lambda}
}
in Planck units. Here $\theta_c$ is the critical value before
backreaction effects dominate, which occurs when
$\delta s(\theta_c)\simeq s_0$ implying\footnote{If the K\"ahler metric for the inflaton depends on more than one saxion, one can extract the value of $\lambda$ from $K_{\theta\theta}^{-1/2}(s^i)\simeq K^{-1/2}_{\theta\theta}(s_0^i)+\delta K^{-1/2}_{\theta\theta}(s^i(\theta))$ with $\delta K^{-1/2}_{\theta\theta}(s^i(\theta))\simeq \lambda \theta$ at large field, and all previous formulae apply.} $\theta_c\simeq
s_0/\lambda$.   In
\cite{Baume:2016psm,Klaewer:2016kiy} it was claimed that
$\lambda$ is a flux independent parameter of order one, implying that
the backreaction effects are therefore tied to the Planck mass. If this is true in general, it is a very powerful
statement which indicates a clear obstruction for  having
trans-Planckian field ranges. 

However, the flux independence of
$\lambda$ was only proved \cite{Baume:2016psm} in type IIA flux
compactifications where the inflaton belonged to the closed string
sector. In \cite{Valenzuela:2016yny} a possible loophole involving the
open string sector was pointed out (and examined in more detail in
\cite{Bielleman:2016olv}). There, the parameter $\lambda$ is not
flux-independent anymore but indeed proportional to the mass hierarchy
$M_\Theta/M_{\rm heavy}$. Therefore a mass hierarchy between the inflaton and
the saxions can help to delay the backreaction effects which are not
anymore tied to the Planck mass. However, the incorporation of more
ingredients to the compactification makes the model more difficult to
control, and it is not clear if such a hierarchy can be really
achieved in a fully reliable global compactification. 

It is the purpose of this paper to  continue the investigation of
these models and similar ones, 
in which $\lambda$ can depend on the above mass hierarchy. We will see that in some representative models, by setting
$\lambda$ small, we are inevitably also decreasing the Kaluza-Klein
scale compared to the moduli mass scale, signaling the breakdown
of the effective theory. But before turning to our results, let us
discuss in more detail the relation between the logarithmic scaling
of the field distance, the breakdown of the effective theory and
the Swampland Conjecture.

\subsection{The Swampland Conjecture}
\label{sec_swamp}

It is clear that not all effective quantum field theories can be obtained as
effective theories from string theory. As made more precise in
\cite{Vafa:2005ui}, besides the string landscape there exist a vast swampland of
such theories that cannot be consistently coupled to  quantum gravity.
In \cite{Ooguri:2006in} Ouguri and Vafa formulated this in a  more concise manner.
They provided a couple of conjectured criteria that an effective theory in the
landscape necessarily should satisfy.
The most quantitative criterium was termed the Swampland Conjecture in
\cite{Klaewer:2016kiy} and it says:

\vspace{0.4cm}
\noindent
{\it Swampland Conjecture:}

For any point $p_0$ in the continuous scalar moduli space of a
consistent quantum gravity theory (the landscape), 
there exist other points $p$ at arbitrarily large distance. As the
distance $d(p_0, p)$ diverges,  an infinite tower of states
exponentially light in the distance appears, 
meaning that the mass scale of the tower varies as
\eq{
\label{swamp_mass}
M \sim  M_0\, e^{-\alpha d(p_0,p)}\,.
} 
Thus, the number of states in the tower which are below any finite mass
scale diverges as $d\to\infty$.

\vspace{0.4cm}
Here, the distance is measured with the metric on the moduli space.
Moreover,  $\alpha$ is a still undetermined  parameter that specifies when 
this behavior sets in, namely beyond $d(p_0,p)\sim \alpha^{-1}$ the
exponential drop-off becomes essential. Infinitely many states
becoming light beyond a certain distance in field space indicates that
the quantum gravity theory valid at the point $p_0$ only has
a finite range $d_{\rm c}$ of validity in the scalar moduli space. As a
consequence any physics that we might derive for larger values
$d>d_{\rm c}$ cannot be trusted.

In this formulation, the flat axion moduli space is assumed to be
compact and the logarithmic behavior is expected to hold rather for the saxions.
Therefore, it is not immediately clear how this conjecture is related to the
question of realizing large field inflation in string theory. How this
proceeds has  been  suggested in \cite{Baume:2016psm,Klaewer:2016kiy} and will also be demonstrated
in the very explicit   prototype models to be discussed in sections
\ref{sec_modela} and \ref{sec_modelb}. Let us already sketch here, how this works.

Say one has managed to stabilize the moduli such that there is only a single
light axion $\Theta$ with mass $M_\Theta$ and a set of heavy other
moduli stabilized at $M_{\rm heavy}$. 
Then,  after integrating out the heavy moduli one can derive an effective 
polynomial potential $V_{\rm eff}(\theta)$ for the light axion, potentially supporting large
field inflation. However, this picture is a bit too naive as we are
interested in field excursion of $\theta$ that are trans-Planckian.
As explained in the previous section, for very large $\theta$ one has to take the backreaction of the 
rolling inflaton onto the other moduli into account. The critical value in proper field space where
this behavior becomes essential is $\Theta_{\rm c}\sim 1/\lambda$ (see eq.\eqref{thetac}).
As discussed above, for field excursions beyond this value, the backreaction causes
the following relation between the proper field distance and $\theta$
\eq{
\label{swampc}
               \Theta={1\over \lambda}\log \left(\theta \right)\,.
}
Therefore, e.g. KK-modes whose mass scales like $M_{\rm KK}\sim
s(\theta)^{-n}\sim \theta^{-n}$
have the scaling $M_{\rm KK}\sim \exp(-n\lambda \Theta)$ with respect to the
proper field distance. This is precisely the behavior stated in the
Swampland Conjecture after identifying
\eq{
             \alpha\sim \lambda \,.
}
Thus, it seems that the original version of the swampland
conjecture can be extended to axion
directions upon taking into account backreaction effects. It is this generalization that we consider in this
paper. Notice that this formulation of the conjecture not only implies a constraint on the field metrics but also on the shape of the scalar potentials coming from string theory, since the backreaction on the saxions is essential to obtain such a logarithmic behaviour at large field.

The essential question now is about the value of $\lambda $. The
original swampland conjecture leaves this open\footnote{For an axion, 
  the WGC implies $f\,S_{\rm inst}\leq 1$ which can  be rewritten in the presence of supersymmetry in terms of the saxionic partner $\varphi$ as
  $\sqrt{g_{\varphi\varphi}}\,\varphi\leq 1$ \cite{Baume:2016psm}. After integration one gets
\eq{ 
&\int \sqrt{g_{\varphi\varphi}} \, d\varphi\leq \int{ 1\over \varphi}\, d\varphi\
\Longleftrightarrow \ \phi\leq  \log\varphi\,,
}
i.e. the  proper field distance grows at best logarithmically as $\phi_{\rm c} \log\varphi$ with $\phi_{\rm c}=\mathcal{O}(1)$.
}.
The set of examples studied in \cite{Baume:2016psm} led the authors to define the so-called {\it Refined Swampland
 Conjecture},  that in addition to the contents of the swampland conjecture
 above states $\alpha=O(1)$. We will see that those examples are only particular cases and that in general one can have
\eq{
\label{massratiolambda}
\Theta_{\rm c}\sim \frac{1}{\lambda}\sim \left(\frac{M_{\rm heavy}}{M_\Theta}\right)^p
}
where $p=0,1$ depending on the model under consideration. In particular, the models in \cite{Baume:2016psm} satisfy $p=0$, while $p=1$ corresponds to the loopholes in \cite{Valenzuela:2016yny,Bielleman:2016olv}. For the latter class of models, if one can manage
to dynamically freeze the moduli such that $\lambda< O(1/10)$, then
one has control over the effective theory for the required $N_e=60$
e-foldings. However we will see that for $\lambda\ll 1$ there are other reasons beyond the exponential
drop-off, why the effective theory fails.

\section{Closed string  models}
\label{sec_modela}

In this section, we revisit  a simple  prototype model
\cite{Blumenhagen:2015kja,Blumenhagen:2015qda} 
of closed string moduli
stabilization and  analyze its  relation to the 
Swampland Conjecture and how this restricts the potential to provide a
controllable string (inspired) model of  F-term axion
monodromy   inflation. 

In \cite{Blumenhagen:2015qda} it was found that the considered single field
inflationary models
with a parametrically light axion fail to also preserve the required
hierarchy of mass scales, thus spoiling parametric control over the
employed effective action. This perfectly matches with the results found in \cite{Baume:2016psm,Valenzuela:2016yny} for their IIA counterpartners. Within the closed string sector of IIA flux compactifications with RR and NS fluxes, it is not possible to get the mass hierarchy required to suppress backreaction, implying that one always get a flux-independent $\lambda\sim \mathcal{O}(1)$. Therefore, we do not expect these closed string IIB models to work either. However, they are a perfect playground to exemplify the backreaction problems and the relation to the Swampland Conjecture. Therefore, instead of analysis an exhaustive list of elaborated models, we will choose the simplest one and discuss the problems arising when trying to drive inflation in the regime $\Theta>\Theta_c$.

Let us emphasize that, in this paper, our focus is on analytically
solvable models,  where in order to be able to compute also the
string and KK-scales, all relevant moduli are included.
It is clear that e.g. the string and the KK-scales are only
dynamically fixed when we include the axio-dilaton as
well as the K\"ahler moduli as dynamical fields.

For the presented representative  examples, we focus on the
parametric dependence of certain relevant quantities in terms
of the background fluxes. Our philosophy is that parametric control is
essential to claim that certain mass hierarchies can be naturally
achieved. Just an accidental, model dependent numerical factor of
e.g. order $O(1)-O(10^2)$ is not sufficient and is certainly not
related to general arguments from quantum gravity.

\subsection{Moduli stabilization and non-geometric fluxes}
\label{sec_fluxes}

Before analyzing concrete models for axion monodromy inflation in
detail,  let us briefly review the necessary concepts of closed string
moduli
stabilization with  various fluxes in type IIB orientifold compactifications.
Later we will not just consider moduli coming from the closed string sector, but are furthermore taking open string moduli into account as they might provide an independent source for inflation. 
Let us postpone the discussion of open string moduli stabilization to
section \ref{sec_OpenModStab}.

We start with compactifying type IIB string theory on orientifolds of Calabi-Yau threefolds $\mathcal M$, which are equipped with a holomorphic three-form $\Omega_3$.
The orientifold projection $\Omega_{\rm P} (-1)^{F_{\rm L}} \sigma$ contains,
besides the world-sheet parity operator $\Omega_{\rm P}$ and the left-moving fermion
number $F_{\rm L}$,
a holomorphic involution $\sigma:{\cal M}\to {\cal M}$. We choose the latter to act on the 
K\"ahler form $J$ and the holomorphic $(3,0)$-form $\Omega_3$ of the Calabi-Yau three-fold $\mathcal M$ as
\eq{
  \label{op_01}
        \sigma^*: J\to +J\,,\hspace{50pt} \sigma^*:\Omega_3\to-\Omega_3\,.
}
The fixed loci of this involution correspond to O$7$- and O$3$-planes, which 
in general require the presence of D$7$- and D$3$-branes  to satisfy the tadpole cancellation
conditions.
The holomorphic involution $\sigma$ of the orientifold projection splits the cohomology into even and odd parts
\eq{
 H^{p,q}(\mathcal M) = H^{p,q}_+(\mathcal M) \oplus H^{p,q}_-(\mathcal M) \,,
 \hspace{50pt}
 h^{p,q} = h^{p,q}_+ + h^{p,q}_-\,.
}
Reducing the ten-dimensional bosonic field content of type IIB string theory on the Calabi-Yau threefold $\mathcal M$ and taking the orientifold projection into account leads to numerous massless moduli in the effective four-dimensional supergravity theory.

The closed string moduli relevant for later constructions  are summarized in table \ref{table_moduli}, where
the convention was chosen such that the imaginary parts of the moduli
correspond to axions\footnote{The full definition of the K\"ahler moduli $T_{\alpha}$ is 
given by
\eq{
\label{defkaehler}
    T_\alpha=\frac{1}{2}\op\kappa_{\alpha\beta\gamma} t^\beta t^\gamma
   +i\left(\rho_\alpha-\frac{1}{2}\op\kappa_{\alpha a b} c^a b^b\right)
  -\frac{1}{4} \op e^\phi  \kappa_{\alpha a b} {G}^a (G+\ov G)^b \,,
}
where $\kappa_{\alpha\beta\gamma}$ denote the triple intersection numbers.}.

\begin{table}[ht]
\centering
\renewcommand{\arraystretch}{1.3}
\begin{tabular}{@{}cl@{\hspace{1pt}}lc@{}}
  \toprule
   number & \multicolumn{2}{c}{modulus} &  name \\
  \hline
  $1$ & $S$&$=g_s^{-1}-i\op C_0$ & axio-dilaton \\
  $h^{2,1}_- (\mathcal M)$ & $U^i$&$=u^i+i\op v^i$ & complex structure\\
 $h^{1,1}_+ (\mathcal M)$ & $T_\alpha$&$=\tau_\alpha+ i \op \rho_\alpha+\ldots$ & K\"ahler\\

 $h^{1,1}_-(\mathcal M)$ & $G^a$&$=S\op b^a+i\op c^a$ & axionic odd\\
\bottomrule
     \end{tabular} 
     \caption{\small Closed string moduli in type IIB orientifold compactifications.}
      \label{table_moduli}
\end{table}

Note that in the following we have redefined the axio-dilaton as $S =
s + i \, c$.  Moduli are stabilized by turning on non-trivial background
fluxes generating a scalar potential for the moduli, see for instance
the review \cite{Blumenhagen:2006ci}. 
Here we will not just focus on R-R and NS-NS three-form fluxes, but
supplementary make use of 
geometric and non-geometric fluxes. For more details we refer to 
\cite{Blumenhagen:2015kja} as well as references therein.

As already mentioned in the introduction, for single field inflation
one needs to achieve a considerable 
mass hierarchy between the inflaton and the other moduli. The KKLT and
Large Volume Scenarios (LVS) \cite{Kachru:2003aw,Balasubramanian:2005zx}
incorporate small non-perturbative effects to fix certain saxionic
K\"ahler moduli, which makes it unnatural to obtain  a mass hierarchy
with the axionic inflaton stabilized at tree-level. 
Therefore, it is  more natural to fix all moduli already at
tree-level by 
employing geometric and non-geometric fluxes for the stabilization of the 
K\"ahler moduli.
Such  fluxes appear  in the context of $\mathcal{N}=2$ gauged
supergravity and double field theory.  However, for completeness, we
will also analyze models within the framework of KKLT and LVS without non-geometric fluxes in section \ref{sec_KKLTLVS}.

In addition to the usual R-R and NS-NS three-form fluxes $\mathfrak F
= \langle dC_2\rangle $ and $H = \langle dB_2\rangle $ there are the
geometric flux $F^I_{\ J K}$ and the non-geometric fluxes $Q_I^{\ J
  K}$ and $R^{IJK}$. Including these new fluxes, the Gukov-Vafa-Witten
superpotential \cite{Gukov:1999ya} can be 
extended in the following compact way \cite{Benmachiche:2006df,Shelton:2005cf}
\eq{
  \label{s_pot_01}
  W = \int_{\mathcal M} \Bigl[ \mathfrak F + \mathcal D \, \Phi^{\rm ev}_{\rm c} \Bigr]_{3} \wedge \Omega_3
  \,,
}
with the complex multi-form $\Phi^{\rm ev}_{\rm c} = i\op S -i\op G^a \omega_a -i \op T_{\alpha} \op\tilde\omega^{\alpha}$and the cohomology bases $\{\omega_a \} \in H^{1,1} ( \mathcal M )$ and $\{\tilde \omega^\alpha \} \in H^{2,2} ( \mathcal M )$. The twisted differential $\mathcal D$ is defined by
\eq{
  \label{d_operator_01}
  \mathcal D = d - H\wedge\: - F\circ\: - Q\bullet\: - R\,\llcorner \,,
}
where the operators appearing in \eqref{d_operator_01} implement the mapping
\eq{
  \renewcommand{\arraystretch}{1.2}
  \arraycolsep3pt
  \begin{array}{l@{\hspace{7pt}}c@{\hspace{1pt}}lcl l@{\hspace{7pt}}c@{\hspace{1pt}}lcl}
  H\,\wedge & :& \mbox{$p$-form} &\to& \mbox{$(p+3)$-form} \,,
  \qquad
  F\,\circ & :& \mbox{$p$-form} &\to& \mbox{$(p+1)$-form} \,,\\
  Q\,\bullet & :& \mbox{$p$-form} &\to& \mbox{$(p-1)$-form} \,,
  \qquad
  R\,\llcorner & :& \mbox{$p$-form} &\to& \mbox{$(p-3)$-form} \,.  
  \end{array}
}
One can be more specific about the action of $\mathcal D$ after introducing a symplectic basis for the third cohomology $H^3 (\mathcal M)$ of the Calabi-Yau threefold.
Eventually the non-vanishing flux components\footnote{It turns out
  that the purely non-geometric $R^{IJK}$ flux does not appear in the
  superpotential.} can be summarized by:
\vspace{0.2cm}
\eq{
  \renewcommand{\arraystretch}{1.4}
\begin{array}{@{}c@{\hspace{30pt}}c@{\hspace{30pt}}c@{\hspace{30pt}}c@{}}
\toprule
\mathfrak F & H & F & Q\\
\hline
\{ \mathfrak{f}_\lambda, \tilde{\mathfrak{f}}^\lambda \} & \{ h_\lambda, \tilde h^\lambda \} & \{ f_{\lambda a}, \tilde f^\lambda_{\ a} \} & \{ q_\lambda^{\ \alpha}, \tilde{q}^{\lambda \alpha} \} \\
\bottomrule
\end{array}
}

\vspace{0.2cm}
\noindent
where $\lambda = 0, \dots, h_-^{2,1}$ and the indices $a$, $\alpha$ label the moduli $G^a$, $T_\alpha$, respectively.
Let us stress that all these fluxes, coupling to moduli of the closed
string sector, are quantized and may only take integer values.

Introducing the periods $X^\lambda$ and $F_\lambda$ of the
holomorphic three-form $\Omega_3$, the complex
structure moduli are determined by $U^i = - i X^i/X^0$. 
In terms of the periods, the superpotential \eqref{s_pot_01}
simplifies  to
\eq{
\label{thebigW}
   W=&
  -\bigl({\mathfrak f}_\lambda  X^\lambda -\tilde {\mathfrak f}^\lambda  F_\lambda  \bigr) 
  +i\op S \big( h_\lambda  X^\lambda - \tilde h^\lambda  F_\lambda \bigr) \\[4pt]
  &+i\op G^a  \bigl( f_\lambda{}_a   X^\lambda - \tilde f^\lambda{}_a  F_\lambda \bigr) 
  -i\op T_{\alpha}  \bigl( q_\lambda{}^\alpha   X^\lambda - \tilde q^\lambda{}^\alpha  F_\lambda\bigr)\,.
}
Apparently the superpotential depends only linearly on the moduli $S$, $G^a$, $T_\alpha$ and in particular the K\"ahler moduli couple to non-geometric fluxes at tree-level.
Together with the perturbative K\"ahler potential at large volume and small string coupling
\eq{
  \label{k_pot}
      K=-\log\left(-i\int_{\mathcal M} \Omega_3 \wedge \ov{\Omega}_3\right)-\log\bigl(S+\ov S\bigr)
     -2\log {\cal V} \,,
}
where $\mathcal V$ denotes the overall volume of the Calabi-Yau threefold $\mathcal M$ in Einstein frame, the flux-induced F-term scalar potential of the moduli in the four-dimensional supergravity theory is given by
\eq{
  \label{f_pot}
  V_F = e^{K} \Bigl( K^{I\ov J} D_IW D_{\ov J}\ov W - 3 \op\bigl|W\bigr|^2 \Bigr) \,,
} 
with K\"ahler metric $K_{I\ov J} = \partial_I \partial_{\ov J}\op K$ and K\"ahler-covariant derivative $D_I W = \partial_I W + (\partial_I K)\op W$.
In general, this scalar  potential stabilizes all the moduli and
generates flux-dependent mass terms for them.

The NS-NS fluxes also give rise to generalized Bianchi identities and to
Freed-Witten anomaly cancellation conditions. Let us remark that for the examples to be
discussed in this paper, these will all be satisfied. 
Let us finally remark that most  non-geometric type IIB fluxes
considered in this paper would correspond to 
geometric fluxes in the T-dual IIA compactification.

\subsection{Closed string model: C1}
\label{sec_closedmodel}

Let us  revisit the most simple model of tree-level flux induced
moduli stabilization, that only contains the two always present moduli,
the axio-dilaton $S=s+ic$ and the overall volume modulus $T=\tau+i\rho$.
This exactly solvable  example 
already reveals the main problem with achieving large field inflation for
F-term axion monodromy.  It can be thought of as an  isotropic $T^6$ with
frozen complex structure modulus.

\subsubsection{Moduli stabilization, masses and backreaction}

At large values of the saxions $(s,\tau)$, 
the K\"ahler potential at leading order is given by
\eq{
          K=-\log (S+\ov S)-3\log(T+\ov T)\,,
}
and the flux-induced superpotential is chosen to be
\eq{
            W=-i {\mathfrak f}_0 +ih\, S+iq\, T\,.
}
The resulting scalar potential reads
\eq{
\label{scalarpotentialclosed}
V =
\frac{(h s+ {\mathfrak f}_0)^2}{ 16s\op \tau^3} 
- \frac{6 h q s -2 q {\mathfrak f}_0 }{ 16 s\op \tau^2} 
- \frac{5 q^2}{ 48  s\op  \tau}  
+ \frac{\theta^2}{16 s\op \tau^3} 
}
with the linear combination $\theta=hc+q\rho$.
This field will be our inflaton candidate.
There exists a non-supersymmetric, tachyon-free AdS minimum at
\eq{
\label{minval}
  \tau_0={6 \, {\mathfrak f}_0\over 5\op q}\,,\hspace{20pt}
   s_0={{\mathfrak f}_0\over h}\,,\hspace{20pt}
   \theta_0=0\,.
}
The masses for the canonically normalized fields are
\eq{
\label{massemods}
M_{{\rm mod},i}^2=\nu_i \op{h \op q^3\over {\mathfrak f}_0^2} \,,
}
with $\nu\in\{0, 0.43, 0.21,0.78\}$.  The cosmological constant
in the minimum is $V_0=-{25\over 216}{h q^3\over {\mathfrak f}_0^2}$.

\begin{figure}[!h]
 \centering
 \includegraphics[width=0.48\textwidth]{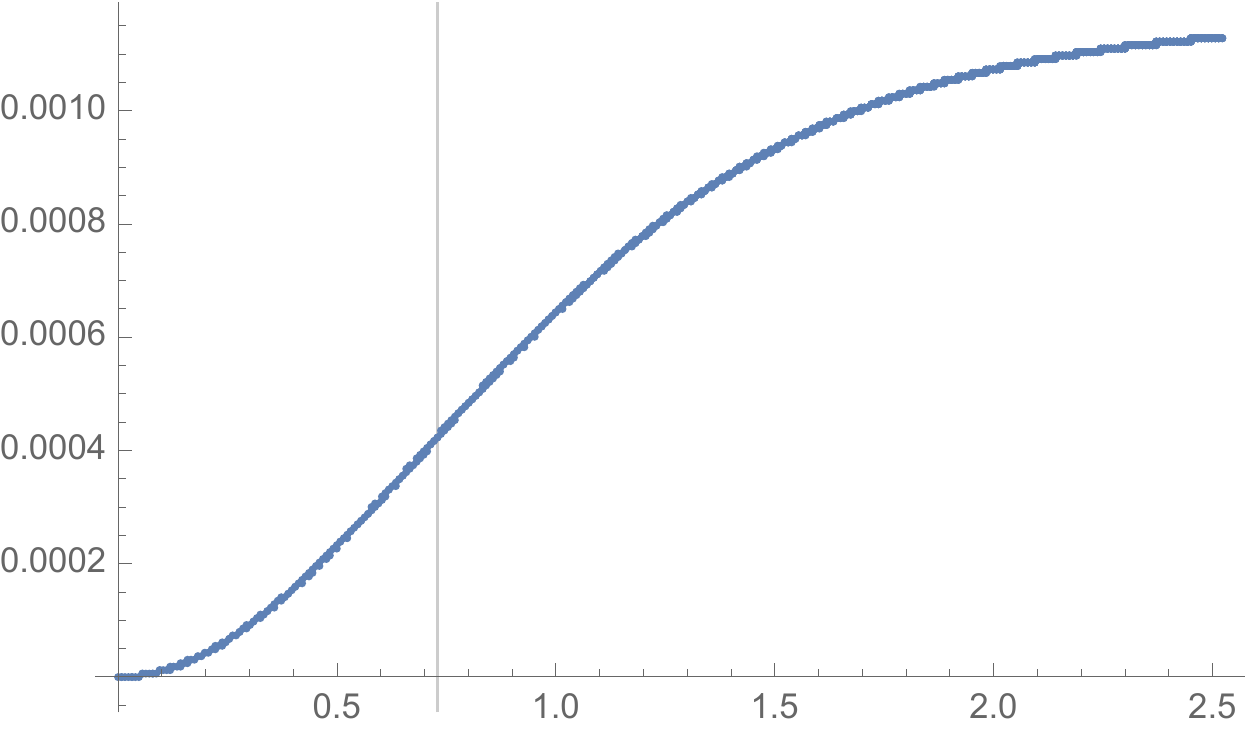}
 \begin{picture}(0,0)
   \put(0,6){$\Theta$}
   \put(-200,124){$V_{\rm back}$}
    \put(-138,-8){$\Theta_{\rm c}$} 
  \end{picture}
\vspace{0.2cm}
\caption{\label{fig_backclosed} The backreacted potential $V_{\rm
    back}(\Theta)$ (after adding a constant uplift) depending on the proper field distance.}
\end{figure}
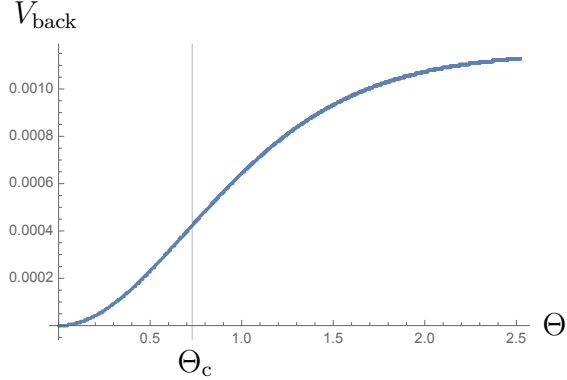

Thus, the mass of the axion $\theta$ is parametrically of the same
order as the masses of the two saxions. Comparing to section \ref{sec_swamp},
this means that $\lambda=O(1)$ and the backreaction should set in 
right at the Planck-scale. Indeed, for field excursions in the
direction $\theta$, the backreaction on the saxions can be exactly
solved and gives
\eq{
\label{backshift}
    \tau_0(\theta)&={3\over 20\op q} \left(  4 \op {\mathfrak f}_0
      + \sqrt{10\op \theta^2  +16\op
        {\mathfrak f}_0^2} \:\right),\\
    s_0(\theta)&={1\over 4\op h}  \,\sqrt{10\op \theta^2 +16\op  {\mathfrak f}_0^2} \, .
}
Looking at the discriminant, it is clear that beyond the critical 
field-value $\theta_c = \sqrt{8\over 5}{\mathfrak f}_0$ the backreaction
becomes substantial. The kinetic term for $\theta$ is
\eq{ 
\label{axkin}
{\mathcal L}_{\rm kin}^{\rm ax} = 
\frac{3}{4(3h^2 s^2 + q^2 \tau^2)} \,{\partial_\mu\theta \partial^\mu \theta}\,,
}
implying that for $\theta<\theta_c$ the canonically normalized
axion is $\Theta={5\over \sqrt{74}} {\theta\over {\mathfrak f}_0}$.
The critical proper field distance is  flux independent
$\Theta_c=\sqrt{20\over 37}\approx 0.73$, i.e.
for the canonically normalized axion the backreaction becomes
substantial right at the Planck-scale. 
The backreacted potential as a function of the proper field distance
is shown in figure \ref{fig_backclosed}. Note that we added a constant
uplift.

\vspace{0.5cm}

\noindent 
It is evident that beyond $\Theta_c$ the potential is not any more of
quadratic form and therefore one cannot realize large field inflation. 
Indeed, in the trans-Planckian regime one finds 
\eq{
{\mathcal L}_{\rm kin}^{\rm ax}={2\over\gamma^2} \left({\partial
    \theta\over \theta}\right)^2\,,
}
with $\gamma=2\sqrt{7\over 5}$. 
The canonically normalized field can be defined as
\eq{
                      \Theta= {2\over \gamma}\log\left({\theta\over 2\theta_c}\right)\,.
}
This is precisely the logarithmic behavior \eqref{swampc} satisfying $\lambda\sim \mathcal{O}(1)$ expected from the Refined Swampland Conjecture.
After assuming a constant uplift by $|V_0|$, the scalar potential reads
\eq{
  \label{back_02}
      V_{\rm back}(\Theta)=|V_0|  \left[1-\left({2\theta_c\over \theta}\right)^2\right]
      =|V_0|\Big[1-e^{-\gamma\op\Theta}\Big]\,.
}
Like the Starobinsky model, $V_{\rm back}$ is a plateau potential for
$\Theta>\Theta_c$. 

Therefore, the strong backreaction led to a
significant flattening of the potential, the initial quadratic
potential of the axion became plateau-like. 
If  $H_{\rm inf}<M_{\rm mod}<M_{\rm KK}$ could be  parametrically guaranteed,
the potential \eqref{back_02}  by itself could
still support inflation 
with a resulting lower value of the  tensor-to-scalar ratio
\eq{
r={8\over (\gamma N_e)^2} \sim O(10^{-3})\,.
}
This looks promising at a first glance, but as we work just at the limit  of having
control,  there are three serious caveats:
\begin{itemize}
\item{In the trans-Planckian regime, the KK-masses show the expected
exponential drop-off
\eq{ 
           M_{\rm KK}\sim {1\over \tau} \sim {q\over {\mathfrak f}_0} \exp\left(-
             {\gamma\over 2}\Theta \right)\,,
}
while the inflationary mass scale $M_{\rm
  inf}=|V_0|^{1\over 4}$ stays constant on the  plateau. Using the relation $V_0=3
M_{\rm pl}^2\, H^2_{\rm inf}$, one finds for the
  ratio
\eq{        {M_{\rm KK}\over H_{\rm inf}}\sim {1\over (q\,
      h)^{1\over 2}} \exp\left(-
             {\gamma\over 2}\Theta_* \right) \,.
}
Thus we parametrically get
$H_{\rm inf}\parag M_{\rm KK}$ so that we are outside the regime of controlling
the effective action.
}
\item{We were assuming here a constant uplift potential, 
  which is however not realistic, as in string theory all known
  potentials drop-off at infinity. The task then is to identify 
   a realistic uplift term that still admits the plateau up to the 
pivot scale before it drops-off towards larger values for the
inflaton. This issue will be addressed below in section \ref{sec_uplift}.}
\item{Since the mass of the inflaton candidate
is of the same scale as the mass of the other moduli, the latter
cannot 
really be integrated out and one has to treat the model in the
framework of multifield inflation. This will affect the trajectory and
the scalar potential along it.}
\end{itemize}

Thus, this example confirms in an analytically deducible way the
statement of the {\it Refined Swampland Conjecture} even for the case
of axionic fields with a shift symmetry. It is the backreaction onto
the saxionic fields that limits the parametrically controllable field
range to be smaller than the Planck-scale. We have also identified  a 
 tower of Kaluza-Klein modes that become exponentially light 
in the trans-Planckian regime. Hence, even Starobinsky-like inflation on a 
sufficiently broad plateau is not under parametric control. 

As we will explain next, to get such a plateau is also challenged from
another perspective, namely 
by considering more realistic (non-constant) uplift terms.
This latter point has  also been observed in \cite{Buchmuller:2015oma}
for a class of models including  instanton contributions, like
for KKLT or the Large Volume Scenario.

\subsubsection{A semi-realistic uplift}
\label{sec_uplift}

So far we were just assuming  a constant uplift. Due to the
backreaction this implied to a constant plateau for $\Theta\to\infty$.
For models with a realistic uplift potential, like $\ov{D3}$ branes
in a warped throat, such a behavior will not happen.
Instead there will be another critical value  $\Theta_{\rm up}$ 
beyond which the uplift term dominates the backreaction.

For the simple closed  string model from section \ref{sec_closedmodel}, it is found that an uplift potential via 
$\ov{D3}$ branes in a warped throat
\eq{
                    V_{\ov{D3}}={\epsilon\over \tau^2}
}
does not work as the full potential $V_F+V_{\ov{D3}}$ does not admit
tachyon-free Minkowski-minima (after fine-tuning of the warp
factor $\epsilon$). In principle, an assumed
uplift potential
\eq{
                    V_{\rm up}={\epsilon\over s}
}
works much better\footnote{We do not know which string theoretic, supersymmetry
  breaking  object can lead to this functional form of an uplift
  potential.}. Here, the full potential  provides a tachyon-free
Minkowski-minimum for the values 
\eq{
\label{minval}
  \tau_0={3 \, {\mathfrak f}_0\over 2\op q}\,,\hspace{20pt}
   s_0={7 \op {\mathfrak f}_0\over 2\op  h}\,,\hspace{20pt}
   \theta_0=0\,, \hspace{20pt}
   \epsilon={2 \op q^3\over 9\op {\mathfrak f}_0}\,.
}
Note that in the perturbative regime $\epsilon$ becomes small.
The  masses for the canonically normalized fields scale in the same
way as in  the non-supersymmetric AdS minimum
\eq{
\label{massemods}
M_{{\rm mod},i}^2=\nu_i \op{h \op q^3\over {\mathfrak f}_0^2} \,,
}
with $\nu\in\{0, 0.55, 0.10,0.87\}$. 

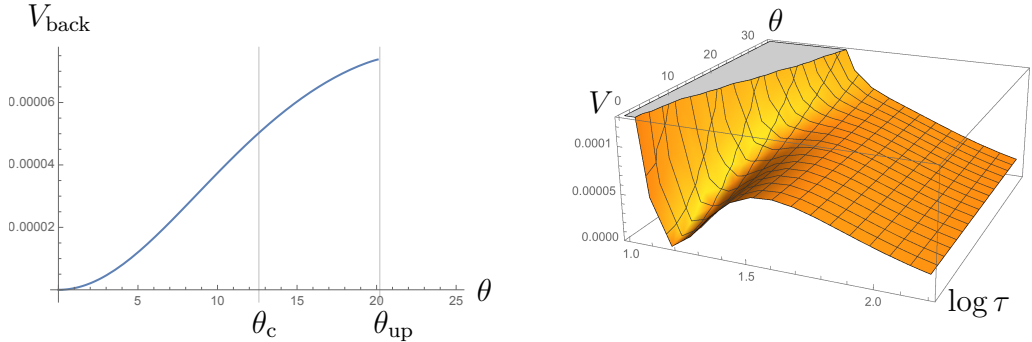
\begin{figure}[!ht]
 \centering
 \subfigure{\includegraphics[width=0.42\textwidth]{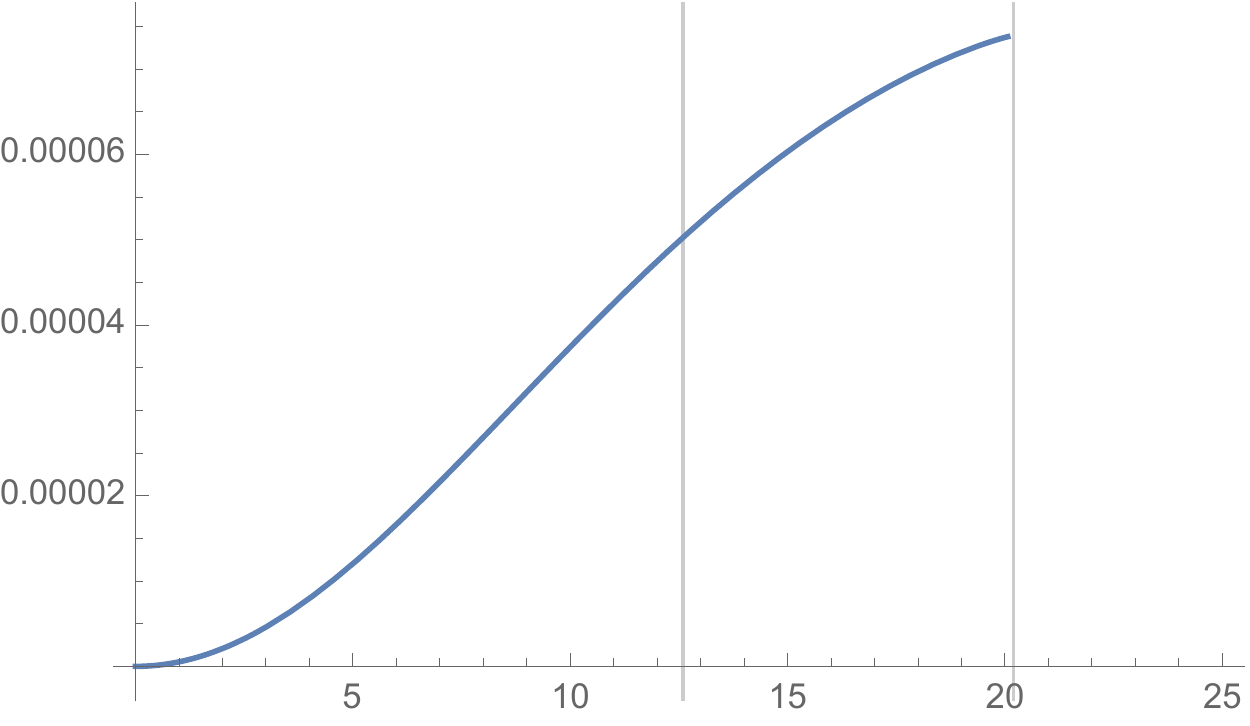}
 \begin{picture}(0,0)
   \put(0,4){$\theta$}
   \put(-170,108){$V_{\rm back}$}
    \put(-85,-8){$\theta_{\rm c}$}
    \put(-39,-8){$\theta_{\rm up}$}
  \end{picture}}
\hspace{1cm}
\subfigure{\includegraphics[width=0.42\textwidth]{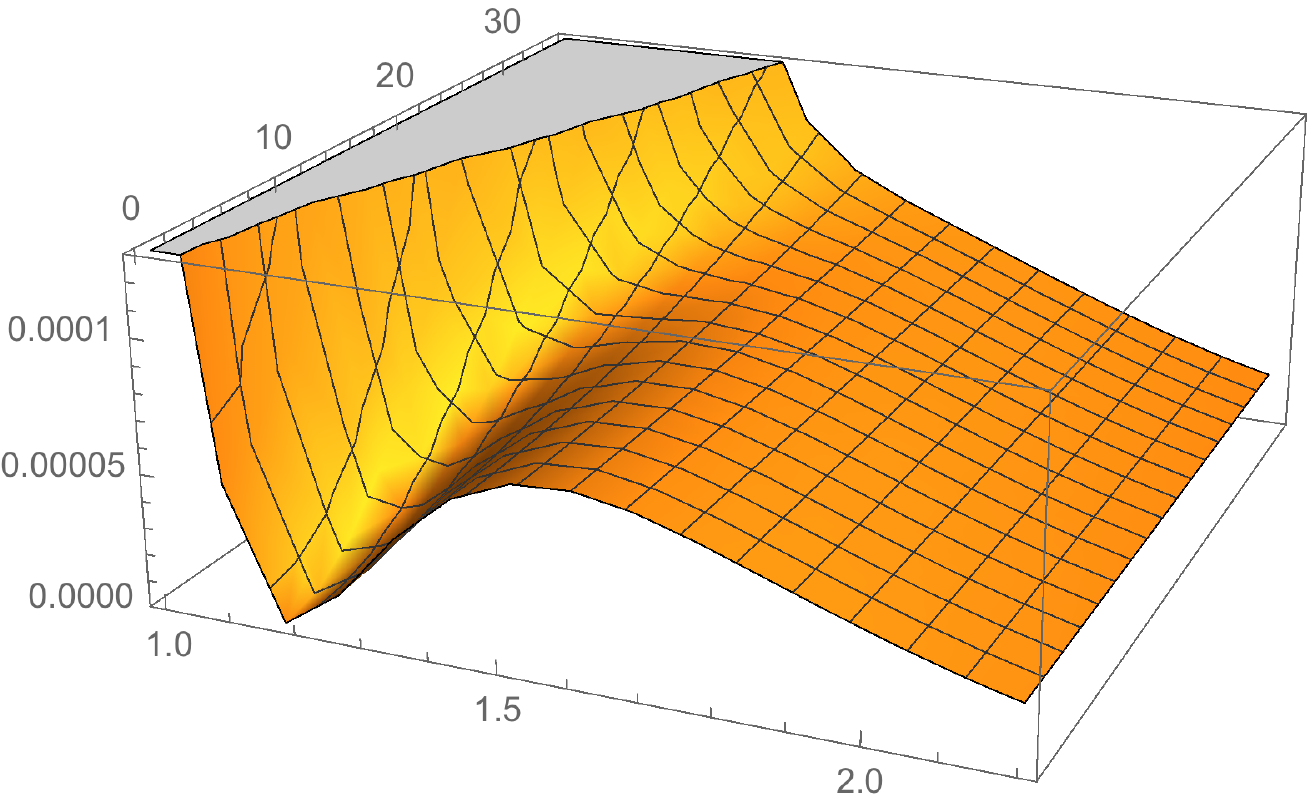}
 \begin{picture}(0,0)
   \put(-35,0){$\log \tau$}
   \put(-171,75){$V$}
    \put(-103,106){$\theta$} 
  \end{picture}
  }
\vspace{0.2cm}
\caption{We plotted on the left the backreacted potential $V_{\rm back}(\theta)$ including the uplift and on the right a slice of  the potential  $V(\theta, s_0(\tau),\tau)$. Both pictures show the destabilization of the inflationary valley.}
\label{fig_uplift}
\end{figure}

When computing the backreaction of a large field excursion of $\theta$
onto the saxions, one finds that the scaling \eqref{backshift} only holds up to a threshold
scale 
\eq{
  \theta_{\rm up}\approx 2\, {\mathfrak f}_0\,,
} 
above which  the uplift term becomes dominant. 
The consequence of this behavior is that for
values $\theta>\theta_{\rm up}$, the local minimum for the saxions
is not present any more, i.e. the valley one is following up comes to
an end at $\theta_{\rm up}$.
This is shown 
for a concrete choice of fluxes in figure \ref{fig_uplift}.
In this example, the critical scale $\theta_{\rm up}$ is between
$\theta_{\rm c}$ (the convex-concave turning scale of the potential)
and the scale where one reaches the top of the plateau. Therefore, for this more
realistic non-constant uplift potential, including the backreaction, one can
never reach the top of the plateau.
Of course, this is just a simple model but, together with the
observations made in \cite{Buchmuller:2015oma}, we think that it
exemplifies another generic obstacle to realize plateau-like large field inflation in string theory.
We will come back to this point when we discuss large field inflation
in KKLT and Large Volume Scenario  in section \ref{sec_KKLTLVS}.

Therefore, it seems clear that one cannot drive inflation in the regime $\Theta>\Theta_c$.
After having familiarized ourselves with the relevant issues that
appear when one wants to realize large field inflation in a controlled
manner,  let us now challenge the {\it Refined Swampland
Conjecture} by trying to follow a recent idea on how one could achieve
a trans-Planckian critical field value $\Theta_c\gg 1$ by introducing open string fields. Notice that we also found a closed string model showing this feature when incorporating an axionic odd $G$ modulus. As it turned out, this model suffers, however, from the same issues which we will describe in the next section about open string moduli.

\section{Open  string  models}
\label{sec_modelb}

The example in the previous section featured $\Theta_c=O(1)$,
providing support for the {\it Refined Swampland Conjecture}.
In this example, $\Theta_c$ was flux independent and we had no chance
to tune it larger. The aim of this central section of this paper is to provide  examples
involving brane deformation moduli that admit an in principle tunable
flux dependent $\Theta_c$.

\subsection{Stabilization of D7-brane moduli}
\label{sec_OpenModStab}
Again, before starting a detailed analysis of models including open string moduli, let us briefly review the necessary conceptual ingredients. 

\subsubsection{D7-brane deformation moduli}

Consider a space-time filling D7-brane with gauge group $U(1)$ wrapping a 4-cycle $C_4$ of the orientifolded Calabi-Yau threefold $\mathcal{M}$. The spectrum of the D7-brane leads to two different types of open string moduli in the 4d effective supergravity theory. On the one hand, there are moduli from deformations transverse to the D7-brane, i.e. D7-brane position moduli, and on the other hand we have Wilson lines of the $U(1)$ gauge field on the 4-cycle $C_4$, see for instance \cite{Jockers:2004yj}.
As shown in \cite{Marchesano:2014iea}, Wilson line moduli are not stabilized by fluxes which makes them unattractive for our setup. For that reason we are exclusively focusing on D7-brane position moduli denoted by
\eq{
	\Phi^I = \varphi^I + i \op \theta^I 
	\qquad {\rm with} \qquad
	I = 1, \dots , h_-^{2,0} (C_4) \, .
	}
If the transverse space of the D7-brane supports 1-cycles, like in a toroidal compactification, the above real fields $\phi^I$ and $\theta^I$ enjoy a shift symmetry. For simplicity we restrict our analysis in the following to the case of a single D7-brane with one complex position modulus $\Phi$.

It is well-known that open string moduli lead to a redefinition of the
holomorphic chiral variables. Whereas Wilson line moduli change the
K\"ahler moduli, the D7-brane position moduli we are employing here,
modify the axio-dilaton $S$
\cite{Jockers:2005zy,Jockers:2004yj,GarciaEtxebarria:2012zm}.
For a D7-brane wrapping a 4-cycle $T^4$ inside $T^6=T^2\times T^4$, the redefinition reads
\eq{
\label{axiodilaton_trafo}
	S \longrightarrow S - \frac{1}{2} \Phi \, \frac{\Phi + \ov \Phi}{U + \ov U} \, ,
	}
with $U$ being the complex structure modulus of the transverse $T^2$. This can be used to determine the K\"ahler potential.
In our prototype models we will compactify on an isotropic six-torus,
whose closed string K\"ahler potential reads
\eq{
	K_{{\rm cl}} = - 3 \log (T + \ov T) - \log (S + \ov S) - 3 \log (U + \ov U) \, .
	}
Taking now also the open string modulus of the D7-brane into account, according to the redefinition of eq. \eqref{axiodilaton_trafo}, one arrives at the K\"ahler potential we will use for our prototype models \cite{Jockers:2004yj}
\eq{
\label{Kpot_open}
	K_{{\rm op}} = - 3 \log (T + \ov T) &- 2 \log (U + \ov U) \\
     &- \log \left[ (S + \ov S)(U + \ov U) - 
	\frac{(\Phi + \ov \Phi)^2}{2} \right]  \, .
	}
It is known that $\alpha'$ corrections from the Dirac-Born-Infeld action of the brane give rise to a non-canonical kinetic term for the inflaton which leads to an additional flattening of the effective scalar potential \cite{Ibanez:2014swa,Bielleman:2016grv}. These corrections will appear as higher derivative corrections to the above K\"ahler potential and can have implications in the determination of the critical value $\Theta_c$. However, since we do not have control over all analogous $\alpha'$ corrections in the closed string sector, we will restric our analysis to leading order in $\alpha'$ in both open and closed string sectors. 

Let us finally specify the superpotential we are working with. It was argued in \cite{Ibanez:2014swa,Arends:2014qca} that D7-brane position moduli give rise to a superpotential of the form
\eq{
	W \supset \mu \, \Phi^2 \, .
	}
Its microscopical origin can be deduced from reducing the DBI and
Chern-Simons actions of the D7-brane or from the T-dual type IIA
description with D6-branes \cite{Carta:2016ynn,Bielleman:2016olv}. Additional motivation
of this superpotential arises from F-theory where complex structure
and D7 position moduli are put on an equal footing. Let us elucidate this
in more detail.

\subsubsection{Superpotential for brane deformations}
\label{app_seven}

Recall that the D7-brane is wrapping a homological  4-cycle $C_4$ in a CY threefold
ambient space $\mathcal{M}$ and is embedded via a map $\iota:C_4\to \mathcal{M}$.
In the perturbative type IIB superstring theory the relevant F-term
potential is (see e.g. \cite{Martucci:2006ij})
\eq{
                    W_o=\int_{\Gamma_5} \Omega_3\wedge (\iota^* B + F)
                    +\Delta W_o
 }  
where $\Gamma$ denotes the 5-chain swept out by pulling the D7-brane
off the orientifold $O(7)$-plane. Moreover,  $\iota^* B$ denotes the 
pull-back of the ambient NS-NS two-form $B$ onto the world-volume
of the D7-brane. The gauge field strength $F$ on the brane
can be expanded into a basis of $H^2(C_4,\mathbb Z)$ and splits 
into two-cocycles that are pull-backs from two-cocycles on $\mathcal{M}$ 
and those whose push-forward to $\mathcal{M}$ is trivial, i.e.
$F=F^\mathcal{M} +\tilde F$. 

Clearly, $\Gamma_5$ depends on the deformation
moduli $\Phi\in H^0(C_4,N_{C_4})= H^{2,0}(C_4,\mathbb Z)$ and the
induced
obstruction appears when by pulling off the brane from the $O7$-plane
a $(0,2)$-component of ${\cal F}=(\iota^* B + F)$ is generated. 
Since the CY ambient space itself does not have any closed $(0,2)$
form, this can only happen if $dB=H\ne 0$ or for the flux components
$\tilde F$  that are cohomologically trivial on $\mathcal{M}$.
In a toroidal set-up, the generation of such an obstruction via a
non-trivial  $H$-flux was demonstrated explicitly in 
\cite{Gomis:2005wc}. The discussion of the $\tilde F$ fluxes appeared
in \cite{Jockers:2005zy} and for toroidal configurations does not
provide a contribution to $W_o$.

Note that in type IIB the co-chain $\iota^* B$  (for $H=dB$) is not 
necessarily quantized as an integer. It was argued in \cite{Arends:2014qca}
that by taking the weak coupling limit of F-theory, an additional term
\eq{
\Delta W_o={i\over 2\pi} \int_\mathcal{M} H\wedge \log\left({P_{D7}\over
  P_{O7}}\right)\, \Omega_3   
}
appears. Here $P_{D7}$ and $P_{O7}$ are polynomials in the coordinates
  on the base that vanish at the location
of the D7-branes and O7-planes, respectively.  In particular, they
depend on the complex structure and brane moduli.
They arise due to the
fact that in F-theory the axio-dilaton is not constant but 
\eq{
            \tau=\tau_0+{i\over 2\pi} \log\left({P_{D7}\over
  P_{O7}}\right)
}
in the orientifold limit.
In F-theory all fluxes reside in $G_4\in H^4(Y,\mathbb Z)$ and are
quantized. Therefore, the extra term $\Delta W_o$ in the type IIB
superpotential can be considered to be necessary for compensating the 
non-quantization of the term involving $\iota^* B$. 

Thus, the naive type IIB superpotential (that treats the brane as a
probe, thus ignoring backreaction effects)
presumably admits non-quantized open string fluxes, whereas in the full F-theory
treatment the quantization of all open and closed string fluxes is 
manifest.

Since the K\"ahler potential that we use is motivated by a single
D7-brane wrapping the isotropic $T^6$, let us lay out what the form of
the superpotential could be.

\subsubsection{Superpotential for D7-brane on a six-torus}

Consider a $T^6=(T^2)^3$ and on each $T^2$ we introduce a
complex structure via $z_a=x_a+i U_a\, y_a$ with $a=1,2,3$. Moreover, we introduce  a
D7-brane wrapping the first two $T^2$ factors. Since this brane does
not contain any 2-cycles that are trivial in the bulk $T^6$, the only
source for a brane superpotential is a non-vanishing $H$-flux.
Such a flux will however generate both a bulk and  a brane superpotential.

Using the conventions and techniques from
\cite{Gomis:2005wc}, let us see what  type of terms can in principle
be generated. Turning  on the general $H_3$ form flux
\eq{
        H= &h_0\, dy_1\wedge dy_2\wedge dy_3 + \\
& h_1\, dx_1\wedge dy_2\wedge dy_3
    + h_2\, dy_1\wedge dx_2\wedge dy_3+h_3\, dy_1\wedge dy_2\wedge dx_3 +\\
& \tilde h_1\, dy_1\wedge  dx_2\wedge dx_3
    + \tilde h_2\, dx_1\wedge dy_2\wedge dx_3+\tilde h_3\, dx_1\wedge dx_2\wedge dy_3 +\\
  &\tilde h_0\, dx_1\wedge dx_2\wedge dx_3
\,,
}
introduces a bulk superpotential 
\eq{
            W_b= \Big(h_0 &-i h_1 U_1 -i h_2 U_2-i  h_3 U_3-\tilde h_1 U_2
            U_3 \\
                  &-\tilde h_2 U_1 U_3-\tilde h_3 U_1 U_2 +i \tilde h_0
            U_1 U_2 U_3\Big)\, iS\,.
}
Here all fluxes are integers and, since the $H$-fluxes do have one leg
on each $T^2$ factor, the Freed-Witten anomaly cancellation condition
$\int_{\rm D7} H=0$ is satisfied.
In order to find the open string superpotential, we restrict the
three-form onto the brane-worldvolume 
\eq{
        B_{\rm D7}= h_0\,  y_3 \, dy_1\wedge dy_2   +\ldots
        + \tilde h_0\, x_3\, dx_1\wedge dx_2\,.
}
Now, we have to check whether this contains a $(0,2)$
component. Indeed, we find
\eq{
     B^{(0,2)}_{\rm D7}= \omega^{(0,2)}\bigg[ &{\partial_S
       W_b\over 2{\rm Re}(U_3)}\,
     (\Phi-\ov\Phi) \\&+ \Big(   -h_3+i \tilde h_1  U_2
     +i \tilde h_2  U_1+ \tilde h_0  U_1 U_2 \Big)\, \Phi \bigg] 
}
where $\Phi=z_3$ and 
\eq{
\omega^{(0,2)}={d\ov z_1\, d\ov z_2\over 4\,{\rm Re}(U_1) {\rm Re}(U_2)} 
}
denotes the $(0,2)$-form on the worldvolume of the D7-brane.
On the supersymmetric locus $\partial_S W=0$ 
the $(0,2)$ component of $B$ depends holomorphically on the brane
position as 
\eq{
     B^{(0,2)}_{\rm D7}= \Big(  -h_3+i \tilde h_1  U_2
     +i \tilde h_2  U_1+ \tilde h_0  U_1 U_2 \Big)\, \Phi\, \omega^{(0,2)} \,.
}  
Therefore, the brane position is frozen at $\Phi=0$.
In the full F-theory picture, where the brane is not
treated as a probe in a supersymmetric bulk,  the bulk/brane superpotential
is expected to read
\eq{
\label{superpotflux}
              W_{\rm tot}=&\, i h_0 S + h_1 U_1 S +  h_2 U_2 S + h_3 (U_3
              S -\Phi^2) -i \tilde h_1 U_2 (U_3 S -\Phi^2) \\
                  &-i \tilde h_2 U_1 (U_3 S -\Phi^2) -i \tilde h_3 U_1
                  U_2 S  - \tilde h_0
            U_1 U_2 (U_3 S -\Phi^2)\,.
}
As we want to deal with the most simple model, we  restrict this to
the isotropic torus. We do this in two steps. First we set all complex
structures to be equal, $U_1=U_2=U_3 \equiv U$. 
Then \eqref{superpotflux} becomes
\eq{
\label{superpotfluxb}
              W_{\rm tot}= i h_0 S &+ (h_1+h_2+h_3) U S  - h_3 \Phi^2
              -i (\tilde h_1 +\tilde h_3+\tilde h_3) U^2 S  \\
          &+i (\tilde h_1 +\tilde h_2) U \Phi^2
                 - \tilde h_0  ( U^3 S -U^2\Phi^2)\,.
}
Still treating the various fluxes as independent parameters, the
coefficients of e.g. the $US$-term and the $C^2$-term could be
disentangled. In the following, we will call this the {\it weakly
  isotropic} torus. In
section \ref{sec_modelO1}, we will present an exactly solvable toy
model of this type. Since it has the advantage of being exactly
solvable, many of the issues about large field excursions can be 
seen very explicitly. 

However, thinking of the isotropic torus as proper Calabi-Yau with
only one complex structure modulus, one would not expect to have
more components of the $H$-flux available than the number
of three cycles, that would be $b_3=4$. This is the reason why for the
 {\it strongly isotropic} torus, we also restrict the fluxes to be
symmetric, i.e. $h_1=h_2=h_3 \equiv \mu_1$, $\tilde h_1=\tilde h_2=\tilde h_3 \equiv \mu_2$ and $\tilde h_0 \equiv \mu_3$. 
In this case the superpotential \eqref{superpotflux} becomes
\eq{
\label{superpotfluxc}
              W_{\rm tot}= i h_0 S &+ \mu_1 (3U S  -  \Phi^2)
              -i \mu_2 (3 U^2 S -2 U \Phi^2)   - \mu_3  ( U^3 S -U^2\Phi^2)\,
}
and a $U^n \Phi^2$ term is always accompanied by a corresponding $U^{n+1}
S$ term. We will also discuss examples of this more realistic type in
section \ref{sec_modelO2}.

\subsubsection{Criteria for models with tunable $\Theta_c$}
\label{sec_crittune}

The purpose of introducing open string fields relies on extending our
analysis to models with a tunable flux-dependent critical value
$\Theta_c$. Then, one might be able to delay the backreaction and the
consequent exponential drop-off of the massive states to a
trans-Planckian value for the inflaton $\Theta_c>1$. As first remarked
in \cite{Valenzuela:2016yny}, this requires the minimum of the
potential to satisfy the following condition: 
\begin{quote}
\it $\Theta_c$ will be tunable if one can set the inflaton mass to zero without destabilizing the other scalars.
\end{quote}
In other words, one needs to engineer a flat direction which is
stabilized by an additional subleading flux $\mu$ in a second
step. The new minimum will correspond then to the old minimum (without
the inflaton) corrected by a term proportional to $\mu$. 
This is precisely the approach that was also followed in \cite{Blumenhagen:2014nba}
and  for the \emph{flux scaling models} considered in
\cite{Blumenhagen:2015kja}. It turns out that the backreacted minima
for the saxions - once we move the inflaton away from its minimum -
take the following schematic form,
\eq{
s=s_0+\delta s(\theta)\ ,\quad \delta s(\phi)\simeq \lambda\,\theta
}
with $\lambda$ depending on the mass hierarchy as $\lambda\sim (
  M_\Theta/M_{\rm heavy})^p$. In the closed string
models of section \ref{sec_modela} and those first analyzed in
\cite{Baume:2016psm}, the above condition is not satisfied since the
value of $s_0$ blows up in the limit $\mu\rightarrow 0$. In those
models, the critical canonical field distance before the logarithmic
behavior dominates is inevitably fixed at $\Theta_c=\lambda^{-1}=O(1)$
in Planck units (or equivalently $p=0$). The inclusion of open string
fields allows us to engineer models with $p=1$ that satisfy the
previous condition.

Let us consider the flux superpotential \eqref{superpotfluxc} of the
effective theory of a D7-brane living in a strongly isotropic torus
derived in the previous section. Every term $\Phi^2$ is accompanied by
a bulk term $SU$. This implies that the only superpotential term for
the dilaton which is independent of $\Phi$ is the linear term
$ih_0 S$. Therefore, we need to have $h_0\neq 0$ in order to stabilize the
dilaton while keeping $\theta={\rm Im} (\Phi)$ massless. We also
assume that there are some RR fluxes stabilizing the complex structure
modulus $U$ and a non-geometric flux stabilizing $T$ via a
superpotential term $i q T$. We are left then with two possibilities:
\begin{itemize}
\item $\mu_1 \neq 0$ and/or $\mu_3 \neq 0$:

As a consequence the superpotential mixes real and imaginary parts of the moduli differently (i.e. even and odd powers of the fields), e.g.
\eq{
	W = i h S + \mu_1 (3 U S - \Phi^2) + \dots
	}
The new minimum cannot be understood as a deformation of the old
minimum proportional to $\mu_1$. In particular, the orthogonal
direction to the axionic combination $\sigma_0=hc_0+q\rho_0$ remains
unfixed in the old minimum and gets a vacuum expectation value in the
new minimum proportional to $\mu_1^{-1}$. This modifies the vevs of
the saxions leading to the same parametric dependence on $\mu_1^{-1}$,
so that we do not recover the old minima when setting $\mu_1=0$. The
strong backreaction then implies $\lambda\sim O(1)$ independently of
the flux choice. A solution comes from adding a term $q_1 UT$ with the
non-geometric flux satisfying $q_1=q\mu_1/h$, which vanishes when
$\mu_1$ goes to zero. In this way, the problematic axionic direction
remains unfixed and the new minimum is simply a deformation of the old
minimum, giving rise to a good candidate for having a flux-tunable $\lambda$. 

\item $\mu_1 = \mu_3 = 0$:

The only possibility to stabilize the open string modulus is now to turn on $\mu_2$, hence
\eq{
	W = i h S + i \mu_2 U (3 U S - 2 \Phi^2) + \dots
	}
This model enters within the class of \emph{flux-scaling models} analyzed in \cite{Blumenhagen:2015kja}. The new minimum can be understood as a deformation of the old minimum which goes to zero when $\mu$ is vanishing. This model is thus a good candidate to obtain a $\lambda$ depending on the flux-tunable mass hierarchy. 
\end{itemize}

For later convenience, we dub the first model with $\mu_1,q_1\neq 0$
as O2 and analyze it further in section \ref{sec_modelO2}. Let us
remark, though, that we get the same conclusions from analyzing the
model with $\mu_2\neq 0$ and we do not include the explicit analysis
simply to avoid cluttering and repetition of results. We will also
analyze an extension of O2 by having both $\mu_1$ and $\mu_3$
non-vanishing. This allows us to discuss an example in which the
$\mu$-parameter entering on $\lambda$ is not a flux integer but an
effective parameter depending also on field vacuum expectation
values. Notice that the other possibility, having both $\mu_1$ and
$\mu_2$ non-vanishing, does not really lead to an effective parameter.
This is due to the relative factor of $i=\sqrt{-1}$ in the superpotential.

In addition, one can also consider the weakly isotropic torus
\eqref{superpotfluxb} which allows us to drop the condition of
having the same flux parameter for the $SU$ and $\Phi^2$ terms. In
this manner we can stabilize the dilaton independently of the inflaton,
without the need of a linear term $ihS$. The new minimum will be a
deformation of the old minimum, yielding a good candidate for having
again a tunable flux-dependent $\lambda$. Due to its computational
simplicity, we will first analyze this model, dubbed as $O1$, in section \ref{sec_modelO1}, and leave the model $O2$ for section \ref{sec_modelO2}. 

Our analysis will  show that, in spite of having in principle a tunable flux
dependent $\lambda$, the flux choice required to delay backreaction
cannot be done without losing parametric control of the effective
theory. In particular, by requiring a mass hierarchy leading to
$\lambda<1$, the  moduli masses become heavier than the Kaluza-Klein scale.

\subsection{Open string model: O1}
\label{sec_modelO1}

Consider now  the  so-called $STU$-model extended by a complex open string
modulus $\Phi$ that parametrizes the transversal deformation of the D7-brane.
Here the four complex moduli are
\eq{
        S=s+ic\,,\qquad T=\tau+i\rho\,, \qquad U=u+iv\,,\qquad
        \Phi=\varphi+i\theta\,
} 
where the imaginary parts are axion-like scalars. At large values of the saxions $(s,\tau,u)$, 
the K\"ahler potential at leading order is given as
\eq{
\label{kaehleropen}
          K=-3\log(T+\ov T)&-2\log (U+\ov U) \\
      &-\log\left[ (S+\ov S)(U+\ov U) -{\textstyle {1\over 2}}
            (\Phi+\ov \Phi)^2\right]\,.
}
As we have seen, the model could  be realized as a D7-brane wrapping a
four-cycle $T^4$ on an isotropic $T^6=(T^2)^3$. 
Now we turn on fluxes to generate the superpotential
\eq{
            W={\mathfrak f}_0+3{\mathfrak f}_2\, U^2 -h\, S\,U-q\, T\, U-\mu\, \Phi^2 \, .
}
Note that for the  {\it strongly isotropic} torus, the fluxes $h$ and $\mu$
would not be independent. Thus, this model only makes sense for
the {\it weakly isotropic} torus and could  therefore still be in the
swampland. Nevertheless, as we will see, it reveals  many interesting 
features and hence is a very good toy model to sharpen our tools. Furthermore, in a more complicated Calabi-Yau, one could aim to disentangle the $h$ and $\mu$ fluxes via additional bilinear couplings of the dilaton to other complex structure moduli that contribute to the first but not to the second one. Therefore, it is a good candidate to exemplify the problems arising even if one manages to get $h\neq \mu$.
Let us mention that this model is
related via mirror-symmetry to a type IIA model with only geometric fluxes
\footnote{Applying three T-dualities
  in the three $x$-directions (of $(T^2)^3$),  one gets a type IIA 
  flux model, where the $D7$ becomes a $D6$-brane and the complex structure
  moduli get exchanged with the K\"ahler moduli. The  K\"ahler potential reads
\eq{
          K=-3\log(U+\ov U)&-2\log (T+\ov T) -\log\left[ (S+\ov S)(T+\ov T) -{\textstyle {1\over 2}}
            (\Phi+\ov \Phi)^2\right]\,.
}
and the superpotential
\eq{
            W={\mathfrak f}_6+3{\mathfrak f}_2\, T^2 -f_0\, S\,T-f_1 \, U\, T-\mu\, \Phi^2\,.
}
Here ${\mathfrak f}_6$ denotes a R-R six-form flux, ${\mathfrak f}_2$ a R-R two-form flux and
$f_i$ geometric fluxes. 
}.

\subsubsection{Moduli stabilization and masses}

This model admits an analytically solvable  non-supersymmetric tachyon-free AdS minimum at
\eq{    s_0&={2^{7\over 4}\cdot 3^{1\over 2}\over 5^{1\over 4}} {({\mathfrak f}_0\,
    {\mathfrak f}_2)^{1\over 2}\over h}\,,\qquad 
    \tau_0={5^{3\over 4}\cdot 3^{1\over 2}\over 2^{1\over 4}} {({\mathfrak f}_0\,
    {\mathfrak f}_2)^{1\over 2}\over q}\\
    u_0&={1\over 10^{1\over 4} \cdot 3^{1\over 2}} \left({{\mathfrak f}_0\over
    {\mathfrak f}_2}\right)^{1\over 2}\,,\qquad \varphi_0=0 \\[0.5cm]
   v_0&=hc_0+q\rho_0=\theta_0=0 \, ,
}
leaving one axionic direction unconstrained.   The value of the scalar
potential in the AdS minimum is
\eq{
                 V_0=-{1\over 120\cdot 3^{1\over 2}\cdot 10^{1\over
                     4}}{h\,q^3 \over {\mathfrak f}_0^{3\over 2}\, {\mathfrak f}_2^{1\over 2}}\,.
}
For the canonically normalized mass-matrix we obtain
\eq{
\label{massesa}
         M^2_{\rm closed}=\nu_i\, {h\, q^3\over {\mathfrak f}_0^{3\over 2}\,
           {\mathfrak f}_2^{1\over 2}}
}
with $\nu\in\{0,0.0001,0.0019,0.0029,0.0117,0.0162\}$ and 
\eq{
\label{massesb}
M^2_{\phi}&=
0.0022\, \left[1 + 14 {\mu\over h} + 24  \left({\mu\over h}\right)^2\right] {h\,q^3\over  {\mathfrak f}_0^{3\over 2}\,
           {\mathfrak f}_2^{1\over 2}}\simeq 0.0022\, {h\,q^3\over  {\mathfrak f}_0^{3\over 2}\,
           {\mathfrak f}_2^{1\over 2}}\\
M^2_{\theta}&= 0.0065\, \mu\, {(3.1623+ 8{\mu\over h}) q^3\over  {\mathfrak f}_0^{3\over 2}\,
           {\mathfrak f}_2^{1\over 2}}\simeq 0.0205\, {\mu\,q^3\over  {\mathfrak f}_0^{3\over 2}\,
           {\mathfrak f}_2^{1\over 2}}
}
where on the right hand side we assumed $\mu/h\ll 1$. 
Therefore, in this regime 
the open string axion $\theta$ is {\it parametrically} lighter than all
the other massive moduli, indeed 
\eq{
     {M_{\rm heavy}\over M_{\Theta}}\sim \sqrt{h\over \mu}=\lambda^{-1}\,.
}
Comparing this to the relation \eqref{massratiolambda} from the
general discussion 
of the Swampland Conjecture, one expects
that $\lambda=\sqrt{\mu/ h}$ is the now flux dependent parameter 
that controls the backreaction of the inflaton onto the other moduli.

\subsubsection{Backreaction}

We can now analyze the model further, in particular to relate to the
general swampland discussion in section \ref{sec_swamp}.

Since this model features a parametrically light axion mass, we expect
that the backreaction in the slow-role regime is also under
control. Let us analyze this in more detail under the assumption
$\lambda\gg 1$. Up to subleading corrections of order ${\cal O}(\lambda^{-2})$, 
the conditions for the backreacted minima can be solved 
\eq{  
\label{back_vevs}
s_0(\theta)&\sim {2^{7\over 4}\, 3^{1\over 2}\over 5^{1\over 4}}
  {({\mathfrak f}_0+\mu\theta^2)^{1\over 2}\,
    {\mathfrak f}_2^{1\over 2}\over h}\,,\qquad
            \tau_0(\theta)\sim {5^{3\over 4}\, 3^{1\over 2}\over 2^{1\over 4}} {({\mathfrak f}_0+\mu\theta^2)^{1\over 2}\,
    {\mathfrak f}_2^{1\over 2}\over q}\\
      u_0(\theta)&\sim {1\over 10^{1\over 4} \, 3^{1\over 2}} \left({{\mathfrak f}_0+\mu\theta^2\over
    {\mathfrak f}_2}\right)^{1\over 2}\,
}
with all other fields sitting in their minimum at zero.
Thus, the critical value of $\theta$ where the backreaction becomes
significant is 
\eq{
\theta_{\rm c}=\sqrt{f_0\over\mu}\,.
}
The kinetic term for the inflaton becomes
\eq{
      {\cal L}^{\rm ax}_{\rm kin}=K_{\Phi\ov\Phi} \,\partial_\mu
      \theta \partial^\mu \theta={1\over 8}\sqrt{5\over 2}\, {h\over f_0+\mu\psi^2}
      \left({\partial \theta}\right)^2
}
so that the critical value for the canonically normalized inflaton
field $\Theta$ is 
\eq{
       \Theta_{\rm c}=\gamma
       \sqrt{h\over f_0}\theta_{\rm c}=\gamma \sqrt{h\over \mu}=\gamma \lambda^{-1}\,
}
with $\gamma={1\over 2} \left({5\over 2}\right)^{1\over 4}=0.63$.
Therefore, from this perspective, for $\lambda\ll 1$ and $\Theta\ll
\Theta_{\rm c}$ the backreaction
can be neglected and one gets the effective potential for the inflaton
(after adding a  constant uplift)
\eq{
     V_{\rm eff}&\simeq \, {\mu h q^3\over f_0^{7\over 2}\, f_2^{1\over
         2}} \left(2 f_0 \theta^2 + \mu \theta^4\right)\simeq \, {\mu h q^3\over f_0^{5\over 2}\, f_2^{1\over
         2}}  \theta^2 \,
     \simeq \, {\mu  q^3\over f_0^{3\over 2}\, f_2^{1\over
         2}}  \Theta^2\,.
    }
Note that the quartic term 
is parametrically
suppressed by a factor $\theta^2/\theta_c^2$ relative to the quadratic one.
Thus, it seems that by parametrically choosing $\Theta_{\rm c}\sim
\lambda^{-1}> 10$ one can achieve a stringy model featuring large field
inflation with a quadratic potential. This is consistent with the
observation already made in \cite{Bielleman:2016olv} for a more complicated, only
numerically treatable open string model (without non-geometric fluxes).

Beyond the critical value,  the kinetic term for the inflaton takes
the form
\eq{
      {\cal L}^{\rm ax}_{\rm kin}={1\over 8}\sqrt{5\over 2}\, {h\over \mu}
      \left({\partial \theta\over \theta}\right)^2
}
so that the canonically normalized inflaton shows the logarithmic behavior
\eq{
      \Theta=\Theta_{\rm c}\, \log\left({\theta\over \theta_{\rm c}}\right)\simeq \frac{1}{\lambda}\,\log\theta \simeq {M_{\rm heavy}\over M_{\Theta}}\,\log\theta\,.
}
Let us mention that, in this regime, the backreacted scalar potential (after constant
uplift) becomes
\eq{
\label{pot_plateau2}
         V_{\rm back}&\simeq 
                    |V_0|\left[
               1-\left({\theta_{\rm c}\over \theta}\right)^3\right]
=|V_0|\left[
               1-\exp\left(-3{\Theta\over \Theta_{\rm c}}\right)\right]\,.
}
Thus, in this large field regime $\Theta\gg \Theta_{\rm c}$ the
backreacted potential is not polynomial but of Starobinsky-like
type.

\subsubsection{Mass scales and the Swampland Conjecture}

From the previous section, the model seems promising to realize
large field inflation with an effective quadratic potential once
we are able to choose the fluxes such that $\Theta_{\rm c}\sim
\lambda^{-1}\gg 1$ and $\Theta<\Theta_{\rm c}$. Thus we need
$h/\mu=O(10^2)$. This could easily be achieved, if the flux
$\mu$ could be tuned much smaller than one.
However, the origin of this flux in F-theory suggests that also
this open string flux is a quantized integer (see section \ref{app_seven}).
In this case, one can only introduce a large flux  $h>O(10^2)$. 

The question is whether such large
fluxes are consistent with the use of the low-energy effective field
theory that we employed for our analysis. 
To see what happens let us consider the various mass scales, like
string scale, Kaluza-Klein scales, heavy moduli masses and the inflaton
mass. 
As mentioned in the beginning of this section, 
we will not be concerned with model dependent numerical prefactors,
but will focus on desired  mass hierarchies that are guaranteed or  spoiled
parametrically.

Thus, up to numerical coefficients, the relevant masses scale in the
following way with the fluxes (recall that we set $M_{\rm pl}=1$):
The string scale is
\eq{
\label{stringscalemin}
                      M^2_{\rm s}\sim { 1\over \tau^{3\over
                          2}\, s^{1\over 2}}\sim {h^{1\over 2}\,
                          q^{3\over 2}\over {\mathfrak f}_0\, {\mathfrak f}_2} \,.
}
Moreover, considering our model as being realized on  the isotropic
$T^6$, we now have {\it two} Kaluza-Klein scales
\eq{
                      M^2_{\rm KK}\sim { 1\over \tau^{2} }\,
                      u^{\pm  1}\,,
}
for $u>1$, yielding  a {\it heavy} and a {\it light} Kaluza-Klein mass
\eq{
           M^2_{\rm KK,h}\sim {q^2\over {\mathfrak f}_0^{1\over 2}\, {\mathfrak f}_2^{3\over 2}} \,,\qquad
           M^2_{\rm KK,l}\sim {q^2\over {\mathfrak f}_0^{3\over 2}\, {\mathfrak f}_2^{1\over
               2}} \,.
}
Recall that the mass of the heavy moduli and the inflaton scaled as
\eq{
         M^2_{\rm mod}\sim  {h\, q^3\over {\mathfrak f}_0^{3\over 2}\,
           {\mathfrak f}_2^{1\over 2}}\,,\qquad
M^2_{\Theta}\sim  {\mu\, q^3\over {\mathfrak f}_0^{3\over 2}\,
           {\mathfrak f}_2^{1\over 2}}\,.
}
Therefore, one gets
\eq{
                  {M^2_{\rm s}\over M^2_{\rm KK,h}}\sim \left( {h
                      {\mathfrak f}_2\over q {\mathfrak f}_0}\right)^{1\over 2}\,.
}
Thus, by choosing the fluxes $\{{\mathfrak f}_0,{\mathfrak f}_2,h,q\}$ all of the same size,
parametrically one can still keep all moduli at the boundary of the perturbative
regime and have the {\it heavy} KK-scale parametrically not bigger than 
the string scale, i.e. $M_{\rm s}\paraeq M_{{\rm KK},h}$.

To relate the mass structure of this model to the Swampland Conjecture, 
reviewed in section \ref{sec_swamp}, we can also evaluate the various
mass-scales in the large field regime. Due to \eqref{back_vevs}, this
means that we just have to change  
\eq{
{\mathfrak f}_0\to {\mathfrak f}_0\left({\theta\over \theta_{\rm c}}\right)^2\to
{\mathfrak f}_0\exp\left(2{\Theta\over \Theta_{\rm c}}\right)
}
so that  the  string scale becomes
\eq{
                      M^2_{\rm s}=  M^2_{{\rm s}}\big\vert_0\,
         \exp\left(-2{\Theta\over \Theta_{\rm c}}\right)\,.
}
Similarly, the  KK-scales in the large field regime are
\eq{
           M^2_{\rm KK,h}= M^2_{\rm KK,h}\big\vert_0\, \exp\left(-{\Theta\over \Theta_{\rm c}}\right)\,,\qquad
           M^2_{\rm KK,l}=M^2_{\rm KK,l}\big\vert_0\,\exp\left(-3{\Theta\over \Theta_{\rm c}}\right) 
}
and for the heavy moduli masses we obtain
\eq{
             M^2_{\rm mod}=M^2_{\rm mod}\big\vert_0\,\exp\left(-3{\Theta\over \Theta_{\rm c}}\right) \,.
}
Therefore, all these mass scales show the expected exponential drop
off \eqref{swamp_mass} at large values in the field space. Thus, for very large values of
$\Theta/\Theta_c$ we have many exponentially light states that invalidate the use
of the low-energy effective action.
For still moderate values of $\Theta/\Theta_c$, one might argue that 
this by itself would not be
disastrous, as long as the order is preserved. However, we also get
\eq{
                  {M^2_{\rm s}\over M^2_{\rm KK,h}}={M^2_{\rm
                      s}\over M^2_{\rm KK,h}} \bigg\vert_0
                 \,  \exp\left(-{\Theta\over \Theta_{\rm c}}\right)
}
which means that  for field excursions  $\Theta/\Theta_c>1$ all  {\it heavy} KK-states 
are heavier than the string scale, i.e. $M_{\rm KK,h}\parag M_{\rm s} $.
This invalidates the usage of 
the low-energy effective supergravity action.

This is all consistent with the Swampland Conjecture. The question now
is whether we also get constraints for the critical value
$\Theta_{\rm c}\sim \lambda^{-1}$.
Can it really be tuned by fluxes to be larger than $M_{\rm pl}$ or do
we find support for the {\it Refined Swampland Conjecture} 
that says $\Theta_{\rm c}$ is close to $M_{\rm pl}$?

For this purpose, let us consider the quotient of the  {\it light} KK-mass and the heavy moduli 
mass
\eq{    
\label{ratio_essentielc}
{M^2_{\rm KK,l}\over M^2_{\rm mod}}\sim {1\over h\, q} \,.
}
This ratio is independent of ${\mathfrak f}_0$ and therefore of $\Theta$ in the
large field regime.
Now, we can distinguish two cases:
\begin{enumerate}
\item{In the case that we could  tune $\lambda$ small by choosing the
open string flux $\mu$ small, there is no problem with the mass hierarchies. As discussed in 
section \ref{app_seven}, this would be in principle possible if one just
considers the naive type IIB form of the open string superpotential.}
\item{However, in the backreacted F-theory picture $\mu$ is quantized.
It is obvious that for large $H$-flux $h$ (i.e. $\lambda\ll 1$) the
ratio \eqref{ratio_essentielc}
is parametrically smaller than one and the moduli masses
are heavier than the KK-mass. This spoils the usage
of an effective four-dimensional effective action for studying
the stabilization of the former massless moduli\footnote{Recall that
  for the strongly isotropic torus, one has $\mu=h$ and therefore
$\Theta_{\rm c}=O(1)$ from the very beginning.}.}        
\end{enumerate}
        
For case 2.  one has  $\lambda=O(1)$ and consequently  $\Theta_{\rm c}=O(1)$.
Thus, we found evidence that the distance in proper field space $\Theta$, where the
logarithmic behavior sets in, is around the Planck-scale and
cannot be much increased without invalidating the effective theory.
In addition, this means that the inflaton cannot be kept 
parametrically lighter than  the other moduli.
Therefore, integrating out the latter first is not a self-consistent approach.
We emphasize that this is precisely what the  {\it Refined Swampland Conjecture} states.

With $\Theta_{\rm c}=O(1)$ for trans-Planckian field excursions one
gets the plateau-like potential \eqref{pot_plateau2}. Analogous to the former
closed string example, for the ratio of the  KK-scale to the 
Hubble scale one finds
\eq{        {M_{{\rm KK,l}}\over H_{\rm inf}}\sim {1\over (q\,
    h)^{1\over 2}} \exp\left(-{3\Theta_*\over 2\Theta_c} \right) \,.
}
We again find the parametric relation
$H_{\rm inf}\parag M_{\rm KK,l}$. Having KK-modes  lighter than the
Hubble scale, spoils the possibility of realizing large field plateau-like
inflation in a controlled way.

\subsection{Open string model: O2}
\label{sec_modelO2}

Let us now consider a model on the strongly isotropic
torus. Unfortunately, it is not exactly solvable, but the intuition 
we gained from the previous examples, allows us to
extract the value of $\lambda$ at least in a perturbative approach.
Here we follow the procedure described in section \ref{sec_crittune} 
and laid out in \cite{Blumenhagen:2014nba,Blumenhagen:2015kja}, i.e.
in a first step we freeze all moduli except the axionic inflaton
candidate. Then we scale these fluxes up and introduce an additional
order one flux to freeze the inflaton. As long as the 
initial values of the moduli are shifted only slightly,
we can integrate them out and determine an effective
potential for the inflaton. This allows us to read off
the ratio of the heavy moduli masses and the inflaton masses.
From the former analysis, we expect that this ratio is directly
related  to $\Theta_c=\lambda^{-1}$, the scale which determine
the backreaction.

\subsubsection{Moduli stabilization and masses}

The model  is defined by the same K\"ahler potential
\eqref{kaehleropen} and the superpotential
\eq{
            W=\Lambda \Big(i {\mathfrak f}_1 U +i\tilde {\mathfrak f}_0\, U^3 + i
            h\, S+ iq\, T\Big)-\mu_1\, (3 U  S -  \Phi^2)-q_1\, 3 U T\,,
}
where $\Lambda$ is a large scaling factor of the four fluxes that, in
the first step, will fix all four saxions and two axionic directions. 
It turns out that the effective approach is only justified if one
choose $hq_1-q \mu_1 =0$, i.e. that only the axionic combination $hc+q\rho$
appears  in the superpotential. Thus, the orthogonal combination will remain
massless. Otherwise, we would not recover the old minimum when setting $\mu_1=0$ and the strong backreaction would imply $\Theta_c\sim \mathcal{O}(1)$ from the very beginning.

In the first step, we set $\mu_1=q_1=0$ and find that there exist a
tachyon-free non-supersymmetric minimum at
\eq{    
\label{fixfirststep}
 s_0&={2^{5\over 4}\cdot 5^{1\over 2}\over 3^{9\over 4}}\, {{\mathfrak
     f}_1^{3\over 2}\over
    h\,\tilde {\mathfrak f}_0^{1\over 2} }\,,\qquad 
\tau_0={5^{1\over 2}\over 2^{3\over 4}\cdot 3^{5\over 4}}\, {{\mathfrak
     f}_1^{3\over 2}\over
   q\, \tilde {\mathfrak f}_0^{1\over 2} }\\
    u_0&={5^{1\over 2}\over 2^{1\over 4}\cdot 3^{3\over 4}} \left({{\mathfrak f}_1\over
    \tilde {\mathfrak f}_0}\right)^{1\over 2}\,,\qquad \varphi_0=0 \\[0.5cm]
   v_0&=hc_0+q\rho_0=0 \, ,
}
leaving one axionic direction unconstrained.  The masses of the
massive moduli are all of the same scale
\eq{
\label{modelo1heavy}
                   M^2_{\rm heavy} \sim { \Lambda^2 \, h \,q^3\,  \tilde{\mathfrak f}_0^{5\over 2}\over {\mathfrak f}_1^{9\over 2} }\,.
}
In the second step we now scale $\Lambda$ up and turn on the small
fluxes $\mu_1$ and $q_1$. Since the axion $\theta={\rm Im}(\Phi)$ only
appears in these extra term in $W$, we expect that it receives a
small mass. In order to estimate it, we integrate out the former
stabilized heavy moduli and compute an effective scalar potential for
$\theta$.
In this regime, the canonically normalized mass of the axion $\Theta$
is
\eq{
\label{massinfl02}
               M^2_{\Theta} \sim { \mu_1^2\, q^3\,  \tilde{\mathfrak
                   f}_0^{3\over 2}\over h\, {\mathfrak f}_1^{7\over 2} }\,.
}
so that, for the scale where the backreaction is expected to become
substantial, we obtain
\eq{
              \Theta_{\rm c} \sim  {M_{\rm heavy} \over  M_{\Theta}}\sim  {\Lambda\, h\, \tilde
    {\mathfrak f}_0^{1\over 2}\over \mu_1\,  {\mathfrak f}_1^{1\over 2}}\gg 1\,.
}
This is large for a sufficiently large flux-scaling factor $\Lambda$.
Note that at this
stage, $\Theta_{\rm c}$ is flux dependent and by appropriate choices
can be tuned large.

As in the previous example O1, let us compute the various mass scales.
We obtain for the string scale, the heavy and light KK-scales in the minimum
\eq{
           M^2_{\rm s}\sim {h^{1\over 2}\, q^{3\over 2}\, \tilde {\mathfrak f}_0\over {\mathfrak f}_1^3} \, ,
    \qquad
   M^2_{\rm KK,h} \sim 
                     { q^2\,  \tilde{\mathfrak f}_0^{1\over 2}\over
                       {\mathfrak f}_1^{5\over 2} }\,,\qquad
                    M^2_{\rm KK,l} \sim 
                     { q^2\,  \tilde{\mathfrak f}_0^{3\over 2}\over
                       {\mathfrak f}_1^{7\over 2} }\,.
}
For the ratio of the string and the heavy KK-scale one finds
\eq{
              {M^2_{\rm s}\over M^2_{\rm KK,h} }\sim \left( {h
                  \tilde {\mathfrak f}_0\over q {\mathfrak
                    f}_1}\right)^{1\over 2}\gg 1 \, ,
}
that we require to be parametrically larger than one.
However, the ratio of the light KK-scale and the heavy moduli mass is
given by
\eq{
             {M^2_{\rm KK,l} \over  M^2_{\rm heavy}}\sim 
     {1\over \Lambda^2\, q^2} \left( {q {\mathfrak
                    f}_1}\over h
                  \tilde {\mathfrak f}_0\right)\paral 1
}
which becomes  parametrically small for large $\Lambda$. 
Therefore, even to get all the high scales in the correct order, we
can at best  work at the boundary of parametric control, where all
fluxes are of order $O(1)$.
However, in this case also the critical 
field distance becomes of order one $\Theta_c=O(1)$ for quantized flux $\mu_1$.

The only possible loop-hole could be that $\mu_1$ is not quantized and
can be significantly smaller than one. This will be analyzed next.

\subsubsection{A comment on tuning in the landscape}\label{sec:tuning}

From the discussed examples it is clear that a possible loop-hole is the
assumption about the quantization of the fluxes. Of course, all the
fluxes in the initial superpotential are quantized but, following the
idea of the landscape, one could
imagine that it is a linear combination of
terms that leads to an effective flux $\mu_{\rm eff}$ that eventually appears in
$\Theta_c$. This effective flux could depend, not only on flux integers but, also on vacuum expectation values of other fields. Here we present a model which exemplifies the above idea and discuss the difficulties to get a substantial tuning.

In the framework of the isotropic torus, we can extend the model O2 by
additional flux induced terms\footnote{Applying a T-duality in the
  three $x$-directions, the fluxes $\mu_3$ and $q_3$ become
  non-geometric R-fluxes in type IIA.}
\eq{
            W=\Lambda \Big(i {\mathfrak f}_1 U +i\tilde {\mathfrak f}_0\, U^3 + i
            h\, S+ iq\, T\Big)&-\mu_1\, (3 U  S -  \Phi^2)-q_1\, 3 U T \\
          &+\mu_3\, U^2 (U S -  \Phi^2)+q_3\, U^3 T\,.
}
Again, to control the minimum of the potential we choose the fluxes
such that only the combination $hc+q\theta$ appears in W, i.e.
$hq_1-q \mu_1 =hq_3-q \mu_3=0$. This guarantees that 
all Bianchi identities are satisfied, as well.
Integrating out the heavy moduli, the mass of the inflaton takes the
same form as in \eqref{massinfl02}
\eq{
               M^2_{\Theta} \sim { \mu_{\rm eff}^2\, q^3\,  \tilde{\mathfrak
                   f}_0^{3\over 2}\over h\, {\mathfrak f}_1^{7\over 2} }\,,
}
but with an effective flux parameter
\eq{
               \mu_{\rm eff}^2=\mu_1^2-{5\over 12\sqrt{6}}  
               \left({ {\mathfrak f}_1\over  \tilde{\mathfrak
                   f}_0}\right) \, \mu_1\, \mu_3 +{25\over 54}  \left({{\mathfrak
                   f}_1\over  \tilde{\mathfrak f}_0}\right)^2\, \mu_3^2\,.
}
As mentioned above, this effective parameter is also moduli dependent
and therefore is certainly not an integer. The question is whether in
the perturbative regime ${\mathfrak f}_1> \tilde{\mathfrak f}_0$ (so $s_0,\tau_0>1$), 
the effective flux can be non-zero and significantly smaller than one. 
First, for $\mu_1\ne 0$, the effective flux $\mu_{\rm eff}$ can be
expressed as
\eq{\label{Heff}
             \mu_{\rm eff}^2={63\over 64} \mu_1^2 +{25\over 54} \left({ {\mathfrak f}_1\over  \tilde{\mathfrak
                   f}_0}\right)^{2}\left( \mu_3 -{3 \sqrt{3}\over
                 20 \sqrt{2}} \left({ \tilde {\mathfrak f}_0\over  {\mathfrak
                   f}_1}\right) \mu_1 \right)^2 \ge {63\over 64} \mu_1^2
}
showing that  $\mu_{\rm eff}$ is larger than $63/64\approx 1$.
For $\mu_1=0$, it is also clear that  $\mu_{\rm eff}>25/54$ giving us the
total lower bound for the effective flux. Thus, we conclude that
in this model one cannot substantially tune the effective flux in the
landscape. As a consequence, the critical field distance is still of
order one.

Up to now, we have analyzed all possible models arising from the brane
superpotential \eqref{superpotfluxc} corresponding to a single
D7-brane living on an isotropic torus $T^6$, although the results also
apply to the case of a Calabi-Yau with a single complex structure
modulus. The natural forthcoming step would be to generalize the
previous idea of tuning in the landscape to more elaborated models
including more than one complex structure modulus, with the hope of
getting a more intricate effective flux parameter $\mu_{\rm eff}$ that
can be tuned small.

However, the inclusion of more fields makes it necessary to extend the backreaction analysis to also these new fields and the corresponding KK scales. 
Of course, this issue cannot so easily be addressed
in full generality, but we would like to emphasize a universal
obstacle which seems difficult to overcome even if appealing to
landscape arguments. This universal obstacle is the backreaction
coming from the dilaton field. The best thing one can intend, is to
stabilize the dilaton by inducing mixing terms between the latter and
other complex structure moduli that do not couple to the open string
modulus. In this way, one can hope to decouple the scale of $S$ and
$\Phi$ and delay the backreaction. As pointed out in \cite{Hebecker:2014kva},
this tuning is in principle possible in the context of F-theory, where
the D7 position moduli and the dilaton become part of the complex
structure moduli of the Calabi-Yau four-fold. Let us remark, though,
that this is precisely the mechanism underlying the model O1, in which
in principle one can get a tunable flux-dependent $\lambda$. However,
as we have seen, even in this case the model fails from realizing
large field inflation. The required mass hierarchy cannot be achieved
without getting into trouble with the KK scale. Therefore, we suspect
similar results might hold for more generic models with more than one
complex structure modulus. A more thorough analysis of Calabi-Yau
geometries is surely interesting and deserves more
investigation, so we leave it for 
future work.

\subsection{Models with instanton corrections}
\label{sec_KKLTLVS}

Let us consider now the case of open string models within the
framework of KKLT \cite{Kachru:2003aw} and Large Volume Scenario (LVS)
\cite{Balasubramanian:2005zx}. The inflaton is still a $D7$-brane
position modulus. The 
K\"ahler moduli are not stabilized by non-geometric fluxes, though, but by
non-perturbative effects. These non-perturbative corrections can
arise, for instance, from Euclidean D3-branes or gaugino condensation
of a stack of distant D7-branes.  As in the previous examples, the complex structure and axio-dilaton
moduli will be stabilized by R-R and NS-NS fluxes.

The backreaction of a field excursion of the inflaton onto the complex structure and axio-dilaton moduli
proceeds analogously to the previous section and leads 
to a logarithmic scaling of the proper field distance at large
field. The critical value at which this happens is given by the mass
ratio $M_u/M_\theta$.
In contrast to the previous models, now this value can in principle be tuned large,
because the KK-scale entering \eqref{ratio_essentielc} depends on the K\"ahler modulus whose
stabilization is now disentangled from the stabilization of the
complex structure and axio-dilaton moduli. In fact, in the analysis of the
KKLT and LVS scenarios we will assume a hierarchy of scales
\eq{
                      M_u>M_\tau>M_\theta\,,
}
and  analyze the effective models after integrating out the complex
structure and the axio-dilaton moduli. The question is whether this
effective field theory  also shows the typical control issues that we found for the
previously studied  models. As opposed to the previous flux examples, here the backreaction can
only be determined up to next-to-leading order.  The relevant parameter controlling
when the backreaction of the inflaton field onto the K\"ahler modulus
becomes substantial is  $\theta_c\sim ({M_\tau/M_\theta})^p$. Notice that the saxions
that determine the kinetic term for the inflaton have already been
integrated out. Therefore, one does not see the
logarithmic behavior from the swampland conjecture for very large field excursions.
However, as before, we find  a potential problem that can invalidate the
possibility of large field inflation.

As already observed in \cite{Buchmuller:2015oma,Kallosh:2004yh},  in the presence of a dynamical
uplifting term, the backreaction on the K\"ahler moduli can
destabilize the vacuum. If the relative displacement of the K\"ahler
moduli during inflation is of order
one, the minimum and the maximum of the KKLT potential merge into a saddle
point so that the minimum disappears and the theory decompactifies. This
is the same effect that we also found in section \ref{sec_uplift} for
an uplift  for the closed string model C1. Thus, the trajectory does
not extend into the regime $\theta>\theta_c$. The  question is, then, whether one can parametrically obtain $\theta_c>1$,
i.e. the mass hierarchy between the inflaton and the K\"ahler modulus.
This is an obvious challenge for KKLT and
LVS as the open string modulus is stabilized at tree-level, whereas
K\"ahler moduli are fixed by non-perturbative corrections. 

We also believe that a full
treatment of the backreaction, i.e. including the complex structure
and  axio-dilaton moduli, would also reveal behavior from the swampland conjecture.

\subsubsection{KKLT scenario}

Let us start analyzing the case of KKLT extended by an open string
modulus $\Phi$. The effective theory, once the dilaton and complex
structure moduli are integrated out, is given by
the K\"ahler potential
\eq{
          K=-3\log (T+\ov T) + \frac{(\Phi + \ov \Phi)^2}{2} \,,
}
and the superpotential
\eq{
            W=W_0+\mu\, \Phi^2+A\, e^{-a T}\,.
}
For simplicity we have set $4su =1$ (in
eq. \eqref{Kpot_open}), 
as  one can show that otherwise the
constraints discussed below become  even stronger.
Moreover, we have approximated the K\"ahler potential by assuming a
small real part of the open string modulus ${\rm Re}(\Phi) = \phi$,
which will in fact be stabilized at zero. $W_0$ and the Pfaffian $A$
are  determined in terms of  fluxes and the stabilized values of the
complex structure moduli. In the following we make the assumptions of
KKLT, namely $A=O(1)$ and $W_0\ll 1$. Moreover, we have in mind
that  $\mu$ is quantized so that we will work in the regime $W_0\ll \mu$.

The interplay between large field  inflation and KKLT moduli
stabilization was already analyzed in \cite{Buchmuller:2015oma} and
further examined in  \cite{Bielleman:2016olv}. Here we just borrow
some of the relations derived there.
The supersymmetric AdS minimum of the scalar potential is  at $\Phi=0$
and for a $\tau_0$ satisfying the transcendental relation
\eq{
W_0=-Ae^{-a\tau_0}\left(1+\frac{2 a\tau_0}{3}\right) \, .\label{KKLT}
}
The masses of the K\"ahler modulus and the inflaton $\theta = {\rm Im}(\Phi)$ are given by
\eq{
\label{mthetaKKLT}
M^2_{\tau} = \frac{(a \, W_0)^2}{2  \tau_0}
\ ,\qquad
M^2_\theta=\frac{1}{2 \tau_0^3}\left(\mu^2+\textstyle{{3\over 2}} \mu W_0\right) 
}
where the latter is the sum of  a supersymmetric mass and a
soft mass. 
If the inflaton is displaced away from its minimum, the minimization
condition for the K\"ahler modulus 
changes in such a way that the minimum for $\tau$ becomes $\theta$-dependent with
\eq{\label{tauKKLT}
\tau=\tau_0\left[1+ {1\over 2}\left(\theta\over \theta_c\right)^2 + \ldots\right]
\,,\qquad \theta_c^2={a\tau_0 W_0\over \mu}\,.
}
The backreaction becomes substantial beyond the critical field
distance $\theta_c$.  In the regime of interest $W_0\ll \mu$, the
supersymmetric mass term for $M_\theta$ is dominant so that one gets
the relation
\eq{
             \theta_c=\sqrt{M_\tau\over M_\theta}\,,
}
i.e., as for the  previous examples,  large field inflation is possible
once we parametrically control  the mass ratio ${M_\tau\over M_\theta}>1$.
Let us now analyze  the two possible obstructions mentioned above:

\begin{itemize}
\item {\bf Controlling $\theta_c$}\\
From \eqref{tauKKLT} it is already clear that one cannot get $\theta_c>1$ for $\mu$ quantized and $W_0\ll 1$ (as required in KKLT).
Employing the condition \eqref{KKLT}, we obtain an upper bound
for the critical field distance \footnote{Here we used the fact that the function
  $F(x)= x\,
  e^{-x}\left(1+{2x\over 3}\right)$ is bounded from above by
  $F_{\rm max}=3 \exp\left( -{3\over 2} \right)\sim 0.67\,$.}
\eq{
      \theta_c^2={|A|\over \mu} (a\tau_0)\,
      e^{-a\tau_0}\left(1+\frac{2 a\tau_0}{3}\right)={|A|\over \mu}
      F(a\tau_0)\lesssim
      {|A|\over \mu}\,.
}
Thus, for $A=O(1)$ one can get $\theta_c>1$ only for a parametrically
small value of $\mu$. This was already noticed in \cite{Bielleman:2016olv}. Therefore, the situation is very similar
to the cases studied before, where the K\"ahler moduli were stabilized
via fluxes.
This supports the conjecture that one cannot achieve single large field inflation in a parametrically controlled effective theory.
\item  {\bf Destabilization due to dynamical uplift}\\
As shown in \cite{Buchmuller:2015oma,Bielleman:2016olv,Kallosh:2004yh}, in the presence of an uplift term (which goes to
zero in the decompactification limit) 
the relative displacement of the K\"ahler modulus $\delta\tau/\tau_0$ cannot be made larger than one
since otherwise the AdS minimum and 
the maximum of the potential merge into a saddle point, destabilizing 
the K\"ahler modulus. Thus, around the critical value
$\theta_c$  the inflationary trajectory stops before
reaching the top of the backreacted potential.
\end{itemize}
Let us remark that, unlike in the previous models, there is no problem
related to Kaluza-Klein states becoming light. Indeed, the Kaluza-Klein scale
stays heavier than the rest of the scales as long as 
$W_0\ll 1/(a\sqrt{\tau_0})$, which is  satisfied for large volume.

\subsubsection{Large Volume Scenario}

One could think that the above problems can be avoided by considering
a scheme in which $W_0$ is not necessarily small. This is indeed one of the ideas proposed in \cite{Bielleman:2016olv} to avoid the above control problems. As an example, we now
consider the LVS scenario \cite{Balasubramanian:2005zx} extended by a D7-brane position modulus
$\Phi = \phi + i \theta$. 
The important feature of LVS is that there exists a non-supersymmetric 
AdS minimum in which the leading order $\alpha'$-correction to the K\"ahler potential
is balanced against a non-perturbative correction to the superpotential.
This leads to an exponentially  large overall
volume $\mathcal V$ that  parametrically controls the vacuum against
higher order corrections.

After integrating out the complex structure and axio-dilaton moduli, we get  an effective model for a typical swiss-cheese manifold with large and small K\"ahler moduli $T_b$ and $T_s$, respectively, 
\eq{
	W &= W_0 + A e^{- a T_s} + \mu \Phi^2\,,\\
	K &= - 2 \ln \left[ (T_b +\ov T_b)^\frac{3}{2} - (T_s + \ov T_s)^\frac{3}{2} + \xi \right] + \frac{(\Phi + \ov \Phi)^2}{2} \, .
	}
Here, $\xi$ denotes the usual $\alpha'$-correction term and $W_0$ and
$A$ are treated as effective parameters of order one.
In particular, denoting the overall volume by $\mathcal V \approx
\tau^{3/2}_b$ and the small four-cycle volume as $\tau={\rm Re}(T_s)$,  in the  minimum one gets for their values
\eq{
\mathcal V_0=\frac{3 W_0 \sqrt{\tau_0}}{\sqrt{2} a  A}\,
e^{a \tau_0} \left(1-\frac{3}{4a \tau_0}\right) \, .\label{LVS}
}
The relevant mass scales for this model are given by
\eq{
\label{MassesLVS}
	M_{\mathcal V} \sim {W_0\over  \mathcal V_0^\frac{3}{2}}\, ,
	\qquad
	M_{\tau} \sim {W_0 \over  \mathcal{V}_0}\, ,
	\qquad
	M_{\rm KK} \sim {1\over \mathcal V_0^{\frac{2}{3}}} \, ,
}
where,  compared to ${\cal V}_0$, we have treated the value of
$\tau_0$ as a number of order one.
The requirement of having the small four-cycle K\"ahler modulus
lighter than the Kaluza-Klein scale already imposes  an upper bound for $W_0$,
\eq{
W_0< \mathcal V_0^{1/3}\ .\label{W0KK}
}
The mass of the open string inflaton was derived in \cite{Buchmuller:2015oma} and at
leading order in $1/{\cal V}$ it takes the simple form
\eq{
\label{LVS_V_Infl}
   M^2_\theta\sim \frac{4 \mu^2}{\mathcal V_0^2}\,.
}
The backreaction of an inflaton excursion  onto the K\"ahler moduli
has also been examined in \cite{Buchmuller:2015oma}(eq. (5.21)). At
leading order in $1/{\cal V}$,  it can be expressed as
\eq{
	\mathcal V &= \mathcal V_0 \left[ 1+ O(1)\, {\mu^2 {\cal
          V}_0\over W_0^2} \theta^2 +\ldots \right]\\[0.1cm]
        \mathcal \tau &= \mathcal \tau_0 \left[ 1+ O(1)\, {\mu^2 {\cal
          V}_0\over W_0^2} \theta^2 +\ldots \right]\,,
}
where the order one prefactors include powers of $\tau_0$ and $a$. 
Thus, the critical field distance can be read of as
\eq{
                   \theta_c\sim {W_0 \over \mu {\cal V}_0^{1\over 2}} \sim
                   {M_{\cal V}\over M_{\theta} }\,.
}
and, as usual, is related to the quotient of the masses.
Finally, we are ready to consider the issues we have already encountered for KKLT:
\begin{itemize}
\item {\bf Controlling $\theta_c$}\\
Employing the condition \eqref{W0KK}, we immediately arrive at the
constraint
\eq{
              \theta_c< {1\over \mu {\cal V}_0^{1\over 6}}
}
which for quantized $\mu$ and large volume is parametrically smaller than one.
Only for very small values of $\mu$ with $\mu<{\cal V}^{-{1\over 6}}$
it could exceed the Planck-scale.
Clearly, this problem just reflects the naive expectation that it is
hard to control an inverted  mass hierarchies, i.e. that a
non-perturbative mass term should be larger than a tree-level mass.
\item  {\bf Destabilization due to dynamical uplift}\\
As for the KKLT example, it was found in \cite{Buchmuller:2015oma}
that in the presence of a dynamical uplift, the overall volume gets
destabilized and the theory decompactifies if the energy during
inflation is bigger than the potential barrier. 
This occurs when the displacement of the overall volume field becomes
comparable to the value at the minimum, i.e. at $\theta_c$. Therefore, the trajectory does not extend in the regime $\theta>\theta_c$.
\end{itemize}

\noindent
Hence, LVS does not provide a better framework than KKLT in this regard. We can conclude that for a quantized open string flux $\mu\geq 1$,
the effective KKLT and LVS scenarios for K\"ahler moduli stabilization
feature the similar  control issues that we already saw for the previous
example of tree-level K\"ahler moduli stabilization.

The loophole again comes from considering an effective $\mu$-parameter  depending on other scalars such that it could be tuned small in the landscape. Whether this tuning is indeed possible is still an open question and deserves more investigation. Notice that the difficulties outlined in section \ref{sec:tuning} also apply to these models. Let us also mention that here we are assuming that $W_0,A$ can be disentangled from the mass scale of the complex structure moduli. But it could very well be that in a full fledge global compactification the two parameters controlling the backreaction of complex structure and Kahler moduli are related, which could reveal the behavior from the swampland conjecture at a lower scale than naively expected. Unfortunately, the global 10d action of these scenarios is not known, so we cannot address this issue in more detail for the moment (see though \cite{Ruehle:2017one} for an effective analysis of the effect of field-dependent Pfaffians $A$).

\section{Conclusions}

In this paper we have critically analyzed the possibility of 
realizing large field inflation in the framework of F-term axion
monodromy inflation for  concrete models of string
moduli stabilization. This included revisiting some 
of the earlier attempts \cite{Blumenhagen:2014nba, Blumenhagen:2015kja,Blumenhagen:2015qda,Blumenhagen:2015xpa}, where it was already observed that
once one dials the flux parameters such that a model
of single field inflation arises, one encounters major
obstacles to parametrically control the various mass
hierarchies in the chain
\eq{
\label{hierarch3}
M_{\rm pl}>M_{\rm s} > M_{\rm KK} > M_{\rm mod} > H_{\rm inf} > M_{\Theta} \, .\nonumber
}
It was suggested in \cite{Baume:2016psm,Klaewer:2016kiy}, that these obstacles could be related to
the axionic extension of the Swampland Conjecture, that was proposed to hold 
in a theory of quantum gravity. For large field inflation, the
essential parameter in this conjecture is the critical scale
$\Theta_{\rm c}\sim \lambda^{-1}$, beyond which a field excursion
imply an infinite tower of states to become exponentially light. The purpose of this paper was to continue the investigation initiated in \cite{Baume:2016psm,Klaewer:2016kiy,Valenzuela:2016yny,Bielleman:2016olv} by enlarging the class of models put under the microscope of the Swampland Conjecture.

Discussing both closed and open string models with flux induced
superpotentials in the perturbative
large volume regime, we found 
further evidence for this conjecture to hold in string theory, once
moduli stabilization is taken into account. We explicitly
saw the appearance of   KK-towers of exponentially light states that
could be traced back to the backreaction of a large field excursion on the 
other moduli, leading to the relation for the proper field distance
$\Theta\sim \log(\theta)$. 

\begin{figure}
\centering
\begin{tikzpicture}[xscale=1.2,yscale=1.2]
\draw [<->] (0,4.5) -- (0,0) -- (7.5,0);
\draw[thick,blue] (0,0) to [out=0,in=250] (2.5,2);
\draw[thick,blue] (2.5,2) to [out=70,in=185] (6.5,3.5);
\draw[dashed, thick] (2.5,-0.1)--(2.5,4.3);
\draw[<->] (0.1,3.3) -- (2.4,3.3);
\node[align=center,red] at (5.5,1.3) {Invalidity of effective theory\\ due to Swampland Conjecture};
\node[align=center,above,red] at (1.25,3.3) {sub-Planckian};
\node[align=center,below,red] at (1.25,3.3) {due to RSC};
\node at (5.3,3.7) {Starobinsky-like};
\node at (5.3,3.1) {Inflation};
\node[align=center] at (0.9,1) {Polynomial\\Inflation};
\node at (7.7,0) {$\Theta$};
\node at (2.5,-0.35) {$\Theta_c$};
\node at (0.5,4.4) {$V(\Theta)$};
\end{tikzpicture}
\caption{The plot depicts schematically a typical potential $V (\Theta)$ for an inflaton $\Theta$ achieved via axion monodromy.
Above some critical value $\Theta_c$ the backreaction of the inflaton
onto the other moduli deforms the polynomial potential to a
Starobinsky-like plateau. However, in this regime the effective theory
breaks down according to the swampland conjecture. The refined version
of the conjecture (RSC) sets $\Theta_c \sim 1$, reducing the
controllable  inflaton field range to sub-Planckian distances.}
\label{fig_swamp}
\end{figure}
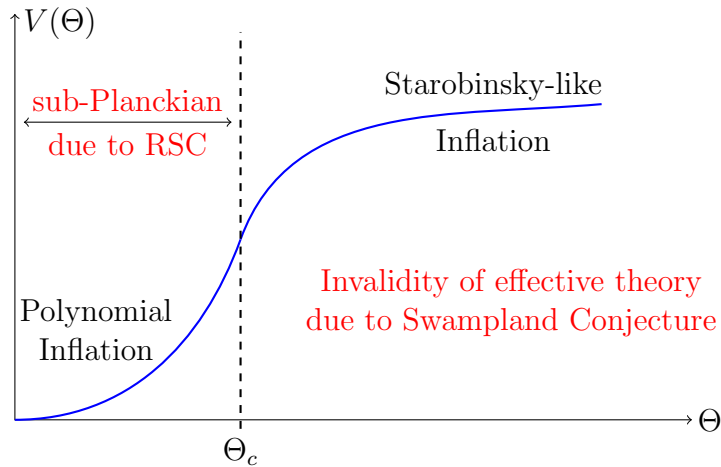

Upon the addition of a constant uplift term, the backreaction of the inflaton onto the other moduli deforms the polynomial potential to a Starobinsky-like plateau above $\Theta>\Theta_c$. However, the appearance of KK-towers invalidates the effective theory in this regime, spoiling inflation. Figure \ref{fig_swamp} illustrates these issues for a typical backreacted axion potential. Furthermore, in the presence of a dynamical uplift which goes to zero at infinite volume, the minimum disappears and the trajectory destabilizes at a scale close to $\Theta_c$,  as already
observed in the framework of large field inflation for KKLT and LVS. The only hope to achieve large field inflation is, thus, obtain a parametrically large value for the critical scale $\Theta_c$.

Whenever the inflationary trajectory can be understood as an original flat direction stabilized by a subleading flux $\mu$, the critical value $\Theta_c$ will depend on the mass hierarchy between the inflaton mass and the heavy moduli masses. If the theory is well behaved in the $\mu\rightarrow 0$ limit and the saxions are not destabilized, one can aim to delay the backreaction effects by increasing the aforementioned mass hierarchy.  We have carefully analyzed effective theories arising from toroidal compactifications of type IIB in which all moduli (including the K\"ahler moduli) are stabilized at tree level by fluxes.
Employing these features for promising models of large field inflation with an open string modulus,
we find that parametric control over the effective supergravity theory eventually required
that the critical scale is just at the Planck-scale, i.e. $\Theta_{\rm
  c}\approx 1$. Consistent with the \emph{Refined Swampland Conjecture}, we could  only achieve a
light axionic  inflaton at the expense of spoiling the validity of the 
four-dimensional  effective action due to a decrease of the Kaluza-Klein scale. 

We also discussed two scenarios (KKLT and LVS)  where the K\"ahler
moduli are not stabilized at tree level by fluxes but by non-perturbative
effects. Similar  control issues arose in the effective
theories after integrating out the complex structure and the axio-dilaton moduli at
a higher scale.\\
 
\vspace{0.1cm}

Thus we conclude: all the previous failing attempts and the concrete string models discussed in this paper support the {\it Refined Swampland Conjecture} \cite{Baume:2016psm,Klaewer:2016kiy}. The take home message is that even if the critical field value $\Theta_c$ at which the effective theory breaks down is in principle a tunable flux-dependent parameter, we find that it cannot be tuned larger than the Planck mass without losing parametric control of the effective theory in all the examples considered so far.

Since our analysis was focusing on obtaining  parametric control, 
we cannot exclude that there might occur accidental coincidences where the
numerical prefactors all work in favor of seemingly generating the
right hierarchy of scales. Though, in all the examples we investigated this does
not happen. 

We think that it is satisfying to see that a general
principle, the {\it Refined Swampland Conjecture}, explains the failure of
all previous attempts to embed the idea of F-term axion monodromy
inflation in the framework of string moduli stabilization. If true, it has huge implications for phenomenology, implying the following result:

\vspace{0.3cm}
\noindent
\begin{center}
\begin{minipage}{12cm}
\emph{In string theory (quantum\ gravity) it is impossible to achieve a
{\it parametrically} controllable model of large (single)
field inflation. The tensor-to-scalar ratio is thus bounded
from above by $r\lessapprox 10^{-3}$.}
\end{minipage}
\end{center}
\vspace{0.3cm}
\noindent

It is a task for the future to gather more evidence for the conjecture or find a model that challenges their implications.
With this in mind, let us mention a few possible loop-holes  that can
trigger further investigation, even though we are not very confident that they will make large field inflation possible. 

It could be that not all fluxes are quantized, as it naively seemed to arise for the type IIB open string superpotential. Alternatively this
could happen after integrating out other more heavy moduli so that 
an effective parameter appears in front of a light modulus in the
superpotential. This is what one usually means by fine-tuning in the landscape. We have analyzed a possible model of this kind within the toroidal framework, without succeed in getting a trans-Planckian field range.
However, whether this can happen in a controlled way in a more generic
F-theory compactification, remains to be seen.  
One related issue is that, introducing more moduli, also means introducing more
KK-scales whose sizes cannot simply be ignored in a honest approach. 
Thus, by referring  too early to the help of a fine-tuning property in the landscape,
the danger is that one  sweeps the dangerous  control issues under
the carpet.

Moreover, we were also restricting the analysis to
 the small string coupling, large radius and
large complex structure regime. It could be that perturbing around
other points in the moduli space works better, even though we expect
that  one faces serious control issues \cite{Blumenhagen:2016bfp}, as well.


\vskip2em
\noindent
\emph{Acknowledgments:} We are very grateful to Joe Conlon, Arthur
Hebecker,  Luis Ib\' a\~nez, Daniel Kl\"awer, Aitor Landete, Fernando Marchesano, Eran Palti, Enrico Pajer and  Timo
Weigand for helpful discussion. IV is supported by  a grant from the Max Planck Society.  Moreover, we would like to thank the
organizers and participants of the workshop ``New Ideas in String
Phenomenology'',  February 14-17, 2017 at  DESY for a stimulating atmosphere,
when part of this work was already presented.



\appendix

\clearpage
\bibliography{references}  
\bibliographystyle{utphys}
\end{document}